\documentclass{article}

\usepackage{amscd,amsmath,amsxtra,amssymb,amsthm,latexsym,amsfonts,color,graphics}

\newtheorem{Thm}{Theorem}[section]
\newtheorem{Lemma}[Thm]{Lemma}
\newtheorem{Prop}[Thm]{Proposition}

\newtheorem{Conjecture}[Thm]{Conjecture}

\newcommand{\R}{{\mathbb R}}
\renewcommand{\O}{{\cal O}}

\newcommand{\C}{{\mathbb C}}
\newcommand{\Z}{{\mathbb Z}}
\newcommand{\N}{{\mathbb N}}

\newcommand{\Tr}{{\rm Tr \ }}

\newcommand{\re}{{\rm Re \ }}
\renewcommand{\Im}{{\rm Im \ }}
\renewcommand{\Re}{{\rm Re \ }}

\newcommand{\bigsigma}{\text{\huge$\sigma$}}

\renewcommand{\Im}{{\rm Im \ }}
\newcommand{\Arg}{{\rm Arg \ }}
\newcommand{\Ln}{{\rm Ln \ }}

\begin{document}

\title{Resurgent Analysis of the Witten Laplacian in One Dimension}

\author{Alexander GETMANENKO \\
{\footnotesize Department of Mathematics, Northwestern University, Evanston IL, U.S.A.;} \\ 
{\footnotesize Max Planck Institute for Mathematics in the Sciences, Leipzig, Germany;} \\
{\footnotesize Institute for the Physics and Mathematics of the Universe, }\\
{\footnotesize The University of Tokyo, 5-1-5 Kashiwanoha, Kashiwa, 277-8568, Japan; }\\ 
{ \tt Alexander.Getmanenko@ipmu.jp}}



\maketitle

\begin{abstract}		

The Witten Laplacian corresponding to a Morse function on the circle is studied using methods of complex WKB and resurgent analysis. It is shown that under certain assumptions the low-lying eigenvalues of the Witten Laplacian are resurgent. 

\end{abstract}

\section{Introduction}

The {\bf goal} of this paper is to study the Witten Laplacian corresponding to a function $f$ on a circle $S^1$ using the methods of resurgent analysis and complex WKB.

The {\bf motivation} of this work comes from three sources. Firstly, there is a number of examples in the literature of computing the spectral asymptotics for different Schr\"odinger equations using the complex WKB method; we are presenting here one more. This example is different in two ways from those described in the literature: namely, the symbol of the Schr\"odinger operator (as an $h$-differential operator) contains more that just a principal part, and instead of considering wave functions on the whole $\R^1$ vanishing at $\pm \infty$, we study periodic boundary conditions. 

Secondly, we believe that methods of resurgent analysis can be helpful in  precise formulation and proof of the conjecture about the products of solutions of different Witten Laplacians that motivate the study of disc instantons and ultimately of the Fukaya category (see ~\cite[\S 5.2]{Fuk}). In his next paper the author hopes to calculate hyperasymptotic expansions of the elements of the Witten complex and to see very explicitly how the product of elements of  Witten complexes corresponding to two different functions is expressed via  contributions from disc instantons.

Thirdly, we want to obtain a connection between WKB and disc instantons in as algebraic way as possible, namely, we want the Planck constant $h$ to be a variable or an asymptotic parameter, not a real number constrained to a small interval. If we succeed in our future research to make an appropriate deformation quantization algebra act on resurgent WKB expansions (perhaps in the spirit of exact deformation quantization, see ~\cite{CM06}), we will come closer to understanding of the Fukaya category in terms of modules over some deformation quantization algebra, which is a project of ~\cite{NT04}. It is this reason that determined our choice of resurgent analysis as a method of investigation.

{\bf Literature review.} (Without any ambition of historical accuracy.) Already in the XIXth century Stokes ~\cite{St1850}  noticed that asymptotic representations of a solution of an ODE in a complex domain  are not uniform with respect to the coordinate, or rather, can be made uniform only in certain regions in the complex plane, called later on {\it Stokes regions.} 

In 1970s these ideas found their application in quantum mechanics (e.g., ~\cite{BB74, BM72}) when the WKB method was extended into the complex domain and asymptotic solutions of the Schr\"odinger equation were sought in terms of ``complex classical trajectories". This naturally involved taking the inverse Laplace transform of the wave function.

Apparent discontinuity in the asymptotic expansions of solutions of differential equations is known as {\it Stokes phenomenon}, and the problem of quantitative calculation of this discontinuity as a {\it connection problem}. See ~\cite{EvFe} for a clean statement and books on special functions of Olver ~\cite{Ol} and Dingle ~\cite{Di} for solutions of connection problems. Malgrange ~\cite{Mal} and Sibuya ~\cite{Si} have applied this to irregular singular ODEs.

Finally, it has been noticed (see \'Ecalle ~\cite{Ec}) that the asymptotic expansions, although divergent, can be Borel resummed. 

These three ideas were nicely (if not completely rigorously) explained and developped in ~\cite{V83} to study  the spectrum of a Schr\"odinger equation precisely enough to catch potentially all  exponentially small contributions. Alternative methods of Helffer and Sj\"ostrand (earlier work) ~\cite{HeSjIV} only give estimates of the instanton effect in the spectrum or calculate it to the leading exponential order as in ~\cite{HeKlNi}; a similar comment applies to the Maslov's complex germ method ~\cite{Mas}. 

A machinery for solutions of the spectral problem in 1D using resurgent analysis is explained and illustrated with many examples in ~\cite{DDP97}. Note also the paper ~\cite{JeZJ04} where some resurgent-analytic study of the supersymmetric double-well Schr\"odinger equation has been done.

It should be noted that in the literature  different kind of asymptotic solutions of differential equations are studied: asymptotics with respect to the coordinate as we move farther away from the singular point of a singular differential equation, and asymptotics with respect to a small parameter $h$ singularly perturbing the equation. Ideas in these two setups are similar, but the details of proofs are, to our knowledge, not directly transferable. We are interested in the latter setup.
    
Let us mention books ~\cite{ShSt,CNP} as systematic texts on resurgent analysis.  

The relationship between the Witten, or Fokker-Planck, Laplacian and the Morse theory was introduced in ~\cite{Wi} and made precise in ~\cite{HeSjIV}. These authors explain that a Morse function on a compact Riemannian manifold $M$ defines Laplacians $P^{(k)}=-h^2 \Delta + |\nabla f|^2 + h({\cal L}_{\nabla f} + {\cal L}^*_{\nabla f})$ on the space of exterior forms $\Omega^k(M)$, define spaces $W^{(k)}\subset \Omega^k(M)$ spanned by eigenfunctions corresponding to its  low-lying, i.e. $O(e^{-c/h})$ for $h\to 0+$, eigenvalues, and prove that $W^{(k)}$ together with the de Rham differential form a complex ({\it the Witten complex}) isomorphic to the Morse complex of the manifold $M$ and the function $f$. 

In the one-dimensional case studied here one takes a standard Riemann metric on the circle, identifies $\Omega^1(S^1)$ with functions on $S^1$ and obtains
$$P^{(0)} = - h^2 \frac{d^2}{dq^2} + (f')^2 - h f'', $$
$$P^{(1)} = - h^2 \frac{d^2}{dq^2} + (f')^2 + h f'' .$$  

Fukaya ~\cite{Fuk} gives intuition to motivate that there should be an $A_\infty$-category structure on Witten complexes corresponding to different Morse functions. In its simplest form the conjecture can be stated as follows: \\ 
\begin{Conjecture} 
Consider functions $f_i$ on a manifold $M$ and Lagrangian submanifolds $L_i = {\rm graph\ }df_i$ in $T^*M$, $i=1,2,3$. Let $p_{ij}$ be an intersection point of $L_i$ and $L_j$ and let $\psi_{ij}$ be a linear combination of eigenfunctions with small enough eigenvalues of the Witten Laplacian on $k$-forms $P^{(k)}$ corresponding to $f_i-f_j$ and localized around $p_{ij}$. Then 
\begin{equation} \langle \psi_{12} \wedge \psi_{23} , \psi_{31} \rangle_{L^2} \ \sim \ \sum  e^{  -\frac{1}{h} \int \omega } \, ||\psi_{12}|| \, ||\psi_{23}|| \, ||\psi_{31}||  \label{eq2} \end{equation}
where the integrals are taken over pseudoholomorphic triangles (``disc instantons'') as above with vertices $p_{ij}$.
\end{Conjecture}

There has been a substantial effort to understand the Fukaya category in terms of microlocal invariants -- note especially ~\cite{NaZa}. The main difficulty of approaching the Fukaya category through semiclassical analysis has been, in our opinion, the fact that on the right hand side of \eqref{eq2} there appear exponentials $e^{-c/h}$ with different $c\in \R$ and that it requires methods of hyperasymptotic analysis to even define what the formula \eqref{eq2} precisely means. We believe that the methods of resurgent analysis, possibly together with related ideas of ~\cite{DiSch}, are a good toolbox to address this difficulty.

It is natural to expect that our WKB approach and the constructible sheaf approach are related by an analog of a Riemann-Hilbert correspondence, yet to be discovered.

{\bf Main result.} The main result of this chapter will be the following:

\begin{Thm} \label{MainResult} Given a generic enough Morse function $f(q)$ in the circle $S^1 \ = \ \R /\Z$ representable as a polynomial in $\sin 2\pi q$ and $\cos 2\pi q$ with $n$ local minima and $n$ local maxima, then the Witten Laplacian
$$ P \ = \  -h^2 \partial^2_q \ + \ [f'(q)]^2 \ - \ hf''(q) $$
with periodic boundary conditions has $n$ resurgent eigenvalues $\le O(h^2)$ corresponding to eigenfunctions that are resurgent in $h$ except possibly for $q\in (f')^{-1}(0)$.
 \end{Thm}


The plan of the proof will be explained in section \ref{PlanOfProof} after we have discussed the generalities of resurgent analysis in section \ref{Preliminaries} and its applications to differential equations in section \ref{ResurgDiffEqu}.

A couple of {\bf comments} are due.

The assumption that $f(q)$ is a generic enough trigonometric polynomial does not seem to be essential to the matter, but simplifies the proofs. We believe that a similar statement holds for any $f$ satisfying $f(q)=f(q+1)$ that can be analytically continued to the whole complex plane. The precise requirement on how generic $f$ should be taken is formulated in terms of the Newton polygon corresponding to the quantization condition, see Section \ref{NewtonPolygon}.

It is the word ``resurgent" in this theorem that constitutes the new result. We are proving that the eigenvalues and eigenfunctions that can be constructed by methods of ~\cite{HeSjIV} and were calculated to the first exponential order by ~\cite{HeKlNi} in fact possess good enough hyperasymptotic expansions. These expansions will be made explicit in our subsequent work.

 The eigenfunctions are claimed to be resurgent with respect to $h$ everywhere except for $q\in (f')^{-1}(0)$. Of course, they will be analytic with respect to $q$ on the whole complex plane, but we do not know if the general theory of resurgent functions (see section \ref{ExistProb}) proves resurgence of eigenfunctions for the values of $q$ where $f'(q)=0$.  This, however, does not seem to be any more that an aesthetic setback.

Resurgent functions are formal objects with respect to $h$, they are in fact classes of functions modulo those that decay faster than any $e^{-c/h}$ for $h\to 0+$ and $c\in\R$, so all equalities between resurgent functions must be understood modulo summands of subexponential decay. We believe that the statements made in this chapter can be promoted to the level of actual functions, but do not address this point.

{\bf The contribution of this paper} is, in our understanding, the following. The Introduction contains a novel research proposal connecting several different ideas in Mathematical Physics. The sections \ref{Preliminaries} and \ref{ResurgDiffEqu} are expository, and their main purpose it to propose terminology that is halfway between those of ~\cite{CNP} and ~\cite{ShSt} and that is relevant for our project. Sections \ref{PlanOfProof}, \ref{FormalWKBSolus}, \ref{StPatternConnProb} contain the standard set-up of the complex WKB method for our specific equation. The section \ref{ConnectionFlae} goes over the derivation of connection formulae given in ~\cite{DP99} to make sure that the arguments given by those authors apply equally well in our case; we hope that we were able to explain this derivation more transparently. Sections \ref{TMQC} and \ref{Solving} contain a new discussion of the structure and solutions of the quantization conditions obtained for the Witten Laplacian. The example in section \ref{example} makes it clear that together with the proof of resurgence of low-lying eigenvalues we have presented a way for effectively calculating them. The method of finding solutions of a quantization condition presented in \ref{ResurgTranscEqu} and of proving that these solutions are resurgent is also new and may be used to improve the level of rigor of other works in resurgent analysis, e.g. ~\cite{DDP97}.


\section{Preliminaries from Resurgent Analysis} \label{Preliminaries}

We need to combine the setup of ~\cite{CNP} (the resurgence is with respect to the semiclassical parameter $h$ rather than the coordinate $q$) and mathematical clarity of ~\cite{ShSt} and therefore have to mix their terminology.  Also, we are somewhat changing their notation to make sure that typographical peculiarities of fonts do not interfere with clear understanding of symbols. We make sure to build the concepts in such a way that sums of exponents such as $e^{-1/h}+e^{-2/h}$ as well as expressions involving logarithms, e.g. $h^k \ln h$, can be treated as resurgent functions.

\subsection{Laplace transform and its inverse. Definition of a resurgent function.}

Morally speaking, we will be studying analytic functions $\varphi(h)$ admitting asymptotic expansions $e^{-c/h}(a_0+a_1 h + a_2 h^2+...)$ for $h\to 0$ and $\arg h$ constrained to lie in an arc $A$ of the circle of directions on the complex plane; respectively, the inverse asymptotic parameter $x=1/h$ will tend to infinity and $\arg x$ will belong to the complex conjugate arc $A^*$.  Such functions can be represented as Laplace transforms of functions ${\bf \Phi}(\xi)$ where the complex variable $\xi$ is Laplace-dual to $x=1/h$ and ${\bf \Phi}(\xi)$ is analytic in $\xi$ for $|\xi|$ large enough and $\arg \xi \in {\check A}$, the copolar arc to $A$. The concept of a resurgent function will be defined by imposing conditions on analytic behavior of ${\bf \Phi}$.

\subsubsection{"Strict" Laplace isomorphism}

For details see ~\cite[Pr\'e I.2]{CNP}.

Let $A$ be a small (i.e. of aperture $<\pi$) arc in the circle of directions $S^1$. Denote by $\check A$ its copolar arc, $\check A = \cup_{\alpha\in A} {\check \alpha}$, where where $\check \alpha$ is the open arc of legth $\pi$ consisting of directions forming an obtuse angle with $\alpha$.

Denote by ${\cal O}^\infty(A)$ the space of sectorial germs at infinity in direction $A$ of holomorphic functions and by ${\cal E}(A)$ the subspace of those that are of exponential type in the direction $A$, i.e. bounded by $e^{K|t|}$ as the complex argument $t$ goes to infinity in the direction $A$.
 
We want to construct the isomorphism
$$ {\cal L} \ : \ {\cal E}({\check A})/{\cal O}(\C)^{exp} \ \longleftrightarrow \ {\cal E}(A^*) \ : \ {\bar{\cal L}}, $$ 
where ${\cal O}(\C)^{exp}$ denotes the space of functions of exponential type in all directions.

{\bf Construction of ${\cal L}$}. Let ${\bf \Phi}$ be a function holomorphic in a sectorial neighborhood $\Omega$ of infinity in direction $\check A$. For any small arc $A'\subset\subset A$ we can choose $\xi_0$ such that $\Omega$ contains a sector $\xi_0 {\check A}'$  with the vertex $\xi_0$ and opening in the direction ${\check A}'$. Define the Laplace transform 
$$ \Phi_\gamma (x) := \int_\gamma e^{-x\xi} {\bf \Phi}(\xi) d\xi $$
with $\gamma=-\partial (\xi_0 {\check A}')$. Then $\Phi_\gamma$ is holomorphic of exponential type in a sectorial neighborhood of infinity in direction $A^*$. Cauchy theorem shows that is ${\bf \Phi}$ is entire of exponential type, then $\Phi_\gamma = 0$, which allows us to define
$$ {\cal L}({\bf \Phi} \ \rm{mod \ } {{\cal O}(\C)^{exp}} ) \ = \ \Phi_\gamma . $$

The construction of ${\bar {\cal L}}={\cal L}^{-1}$ will not be used in this paper.

\subsubsection{"Large" Laplace isomorphism}

The Laplace transform ${\cal L}$ defined in the previous subsection can be applied only to a function ${\bf \Phi}(\xi)$ satisfying a growth condition at infinity. At a price of changing the target space of ${\cal L}$ this restriction can be removed as follows.

For details see ~\cite[Pr\'e I.3]{CNP}.

Let ${\check A}=(\alpha_0,\alpha_1)$ be the copolar of a small arc, where $\alpha_0,\alpha_1\in S^1$ are two directions in the complex plane, and let $\gamma: \R\to \C$ be an endless continuous path. We will say that $\gamma$ is {\bf adapted} to ${\check A}$ if    $\lim_{t\to -\infty} \gamma(t)/|\gamma(t)|\to \alpha_0$, $\lim_{t\to \infty} \gamma(t)/|\gamma(t)|\to \alpha_1$, and if the length of the part of $\gamma$ contained in a ring $\{ z: R \le |z|\le R+1\}$ is bounded by a constant independent of $R$.

Let us now construct for a small arc $A$ two mutually inverse isomorphisms
$$ {\cal L} \ : \ {\cal O}^\infty({\check A})/{\cal O}(\C) \ \longleftrightarrow \ {\cal E}(A^*) / {\cal E}^{-\infty}(A^*) \ : \ {\bar{\cal L}},$$ 
where ${\cal E}^{-\infty}(A^*)$ is the set of sectorial germs at infinity that decay faster than any function of exponential type (cf. ~\cite[Pr\'e I.0]{CNP}).

{\bf Construction of ${\cal L}$}. Let ${\bf \Psi}$ be holomorphic in $\Omega$, a sectorial neighborhood of infinity of direction $\check A$. Let $\gamma$ be a path adapted to ${\check A}$ in $\Omega$. It is known that there is a function ${\bf \Phi}$ bounded on $\gamma$ such that ${\bf \Phi}-{\bf \Psi} \in {\cal O}(\C)$;  define
$$ {\cal L}({\bf \Psi} \rm{\ mod \ } {\cal O}(\C))  \ := \  \int_\gamma e^{-x\xi} {\bf \Phi}(\xi)d\xi \ \mod {\cal E}^{-\infty}(A^*) $$

{\bf Definition.} Any of the functions ${\bf \Psi}(\xi)$ satisfying ${\cal L}({\bf \Psi} \rm{\ mod \ } {\cal O}(\C)) = \psi(x)$ is called a {\bf major} of the function $\psi(x)$.

An equivalence class of functions defined on a subset of $\C$ modulo adding an entire function is also called an {\bf integrality class}.

The map ${\cal L}$ respects linear combinaitions and transforms products of resurgent functions into appropriately defined convolutions of majors.

\subsubsection{Resurgent functions.} \label{ResurFunDef}

Resurgent functions are usually understood to be functions of a large parameter $x$. For brevity we will speak of resurgent functions of $h$ to mean resurgent functions of $1/h$.  

{\bf Definition.} A germ $f(\xi)\in \O_{\xi_0}$  is {\bf endlessly continuable} if for any $L>0$ there is a finite set $\Omega_L\subset \C$ such that $f(\xi)$ has an analytic continuation along any path of length $<L$ avoiding $\Omega_L$.

{\bf Definition.} (cf. ~\cite[p.122]{ShSt}) Let $A\subset S^1\subset \C$ be a small arc and let $A^*$ be obtained from $A$ by complex conjugation. A {\bf resurgent function} $f(x)$ (of the variable $x\to\infty$) in direction $A^*$ is an element of ${\cal E}(A^*)/{\cal E}^{-\infty}$ such that its major $({\bar{\cal L}}f)(\xi)$ is endlessly continuable.

{\bf Remark.} ~\cite{CNP} calls the same kind of object an "extended resurgent function". 

Let (cf. ~\cite[R\'es I]{CNP})  ${\pmb{\cal R}(A)}$ denote the set of endlessly continuable sectorial germs of analytic functions $\Phi(\xi)$ defined in a neighborhood of infinity in the direction $\check A$. Then a resurgent function of $x$ in the direction $A^*$ has a major in ${\pmb{\cal R}}(A)$. 

When we mean a resurgent function $g(h)$ of a variable $h\to 0$, under the correspondence $x=1/h$ the sectorial neighborhood of infinity in the direction $A^*$ becomes a sectorial neighborhood of the origin in the direction $A$, and we will talk about a resurgent function $g(h)$ (for $h\to 0$) in the direction $A$.

\subsubsection{Examples of resurgent functions.}
 
\paragraph{Example 1.} $h^\alpha$ 

The major corresponding to $h^{\nu}$ for $\nu\ne 1,2,3,..$ is 
$\frac{-1}{2i\sin(\pi\nu)} \frac{(-\xi)^{\nu-1}}{\Gamma(\nu)}$, and $\xi^{\nu-1}\frac{\Ln \xi}{2\pi i \Gamma(\nu)}$ for $\nu=1,2,...$.

\paragraph{Example 2.} $\ln h$.

\paragraph{Example 3.}  $\Phi(h):=e^{1/h}$, $\Phi(h):=e^{-1/\sqrt{h}}$. (~\cite[R\'es II.3.4]{CNP})

\paragraph{Example 4.}  $e^{-1/h^2}$ is zero as a resurgent function on the arc  $(-\pi/4,\pi/4)$, but does not give a resurgent function on any larger arc  because there it is no longer bounded by any function $e^{c/h}$ for $c\in \R$.

\subsection{Decomposition theorem for a resurgent function} \label{DecompThm}

Making rigorous sense of a formula of the type
$$\phi(h) \ \sim \ \sum_k e^{-c_k/h}(a_{k,0}+a_{k,1}h+a_{k,2}h^2+...), \ \  \ \  h\to 0,$$
may be done by explicitly writing an estimate of an error that occurs if we truncate the $k$-th power series on the right at the $N_k$-th term. Resurgent analysis offers the following attractive alternative: if $\phi(h)$ is resurgent, then the numbers $c_k$ can be seen as locations of the first sheet singularities of the major ${\bf\Phi}(\xi)$, which is expressed in terms of a decomposition of ${\bf\Phi}(\xi)$ in a formal sum of microfunctions that we are going to discuss now. Microfunctions, or the singularities of ${\bf\Phi}(\xi)$ at $c_k$ are related to power series $a_{k,0}+a_{k,1}h+a_{k,2}h^2+...$ through the concept of Borel summation, see section \ref{BorelSum}.

\subsubsection{Microfunctions and resurgent symbols} \label{MicrofAndResSymb}

{\bf Definition.} (see ~\cite[Pr\'e II.1]{CNP}) A {\bf microfunction} at  $\omega\in\C$ in the direction ${\check A}\subset {\mathbb S}^1$ is the datum of a sectorial germ at $\omega$ in direction $\check A$ modulo holomorphic germs in $\omega$; the set of such microfunctions is denoted by
$${\cal C}^\omega(A) = \check {\cal O}^\omega(A) / {\cal O}_\omega . $$
A microfunction is said to be {\bf resurgent} if it has an endlessly continuable representative. The set of resurgent microfunctions at $\omega$ in direction $\check A$ is denoted ${\pmb{\cal R}}^\omega (A)$ (~\cite[R\'es I.3.0, p.178]{CNP}).

{\bf Definition.} (~\cite[R\'es I.3.3, p.183]{CNP}) A {\bf resurgent symbol} in direction $\check A$ is a collection $\dot\phi = (\phi^\omega \in {\pmb{\cal R}}^\omega (A))_{\omega\in\C}$   such that $\phi^\omega$ is nonzero only for $\omega$ in a discrete subset $\Omega\in\C$ called the {\bf support} of $\dot\phi$, and for any $\alpha\in A$ the set $\C \backslash \Omega\alpha$ is a sectorial neighborhood of infinity in direction $\check A$. 

The set of such resurgent symbols is denoted $\dot{\pmb{\cal R}}(A)$, and resurgent symbols themselves can be written as  $\dot \phi = (\phi^\omega)_{\omega\in \Omega} \in \dot{\pmb {\cal R}}(A)$ or as $\dot \phi = \sum_{\omega\in\Omega} \phi^\omega \in \dot{\pmb {\cal R}}(A)$.

{\bf Definition.} A resurgent symbol is {\bf elementary} if its support $\Omega$ consists of one point. It is {\bf elementary simple} if that point is the origin. 

\subsubsection{Decomposition isomorphism.} \label{DecomposnIsomsm}

The correspondence between resurgent symbols in the direction $A$ and majors of resurgent functions in direction $A$ depends on a {\it resummation direction} $\alpha\in A$, which we will fix once and for all. The direction $\alpha$ can more concretely be thought of as $\arg h$ or as the the direction of the cuts in the $\xi$-plane 
that we are going to draw.

 Let $\dot \phi = (\phi^\omega)_{\omega\in \Omega} \in {\bf \cal R}(A)$ be a resurgent symbol with the singular support $\Omega$, and $\alpha \in A$. Let  $\Omega\alpha=\bigcup_{\omega} \omega\alpha$ be the union of rays in the direction $\alpha$ emanating from the points of $\Omega$:

\begin{figure}[h]\includegraphics{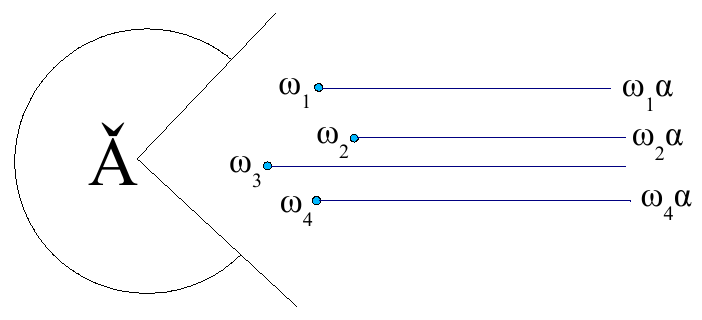} \caption{Singularities of a major and corresponding cuts} \label{ResDRWp8}
\end{figure}

Suppose (~\cite[p.182-183]{CNP}) $\Omega$ is a discrete set (of singularities) in the complement of a sectorial neighborhood of infinity in direction $\check A$, and a holomorphic function ${\bf \Phi}$ is defined on $\C\backslash \Omega \alpha$ and is endlessly continuable. Take  $\omega\in \Omega \alpha$.   Let $D_\omega$ be a small disc centered at $\omega$. Its diameter in the direction $\alpha$ cuts $D_\omega$ into the left and right open half-discs $D^{-}_\omega$ and $D^{+}_\omega$ (or top and bottom if $\alpha=\R_+$). If $D_\omega$ is small enough, the function ${\bf\Phi}|D^{+}_\omega$, resp ${\bf\Phi}|D^{-}_\omega$, can be analytically continued to the whole split disc $D_\omega \backslash \omega \alpha$. Denote by $sing^{\omega}_{\alpha+} {\bf \Phi}$, resp. $sing^{\omega}_{\alpha-} {\bf \Phi}$, the microfunction  at $\omega$ of direction $\check \alpha$ defined by the class modulo ${\cal O}_\omega$ of this analytic continuation. 

\begin{Thm} There is a function ${\bf \Phi}\in {\cal O}(\C\backslash \Omega\alpha)$, endlessly continuable, such that
$$ sing^\omega_{\alpha+} {\bf \Phi} = \left\{ 
\begin{array}{ll} \phi^\omega & \text{if $\omega\in\Omega$} \\ 0 & \text{if not}
\end{array} \right. . $$
In this case we write
$${\bf \Phi} = {\bf s}_{\alpha+}{\dot \phi} $$ and call the inverse map $({\bf s}_{\alpha+})^{-1}$ the {\bf decomposition isomorphism} \end{Thm}

An endlessly continuable function ${\bf s}_{\alpha-}{\dot \phi}$ can be defined analogously. 

The maps ${\bf s}_{\alpha+}$,${\bf s}_{\alpha-}$  respect sums and convolution products (cf. ~\cite[p.185, R\'es I, section 4]{CNP}). 

 If a resurgent function $\varphi(h)=\varphi(h,t)$ and its major ${\bf\Phi}(\xi)={\bf\Phi}(\xi,t)$ depend, say, continuously in some appropriate sense, on an auxiliary parameter $t$, the decompositon into microfuctions $({\bf s}_{\alpha+})^{-1}{\bf \Phi}$ will depend on $t$ ``discontinuously" -- an effect referred to as {\it Stokes phenomenon}

\subsubsection{"Homomorphisme de passage"} \label{HomDePassage}

\begin{figure}
\includegraphics{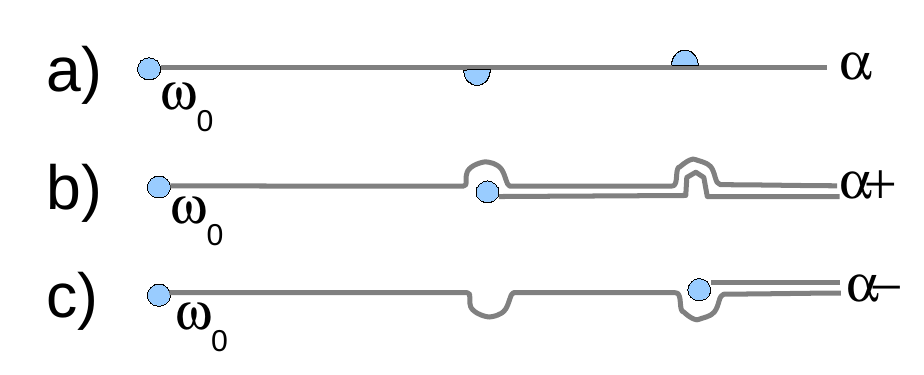} \caption{Difference between ${\bf s}_{\alpha +}$ and ${\bf s}_{\alpha -}$, where $\alpha$ is a positive real direction: a) two additional singularities around the horizontal cut; b) and c) the singularities that appear in the formal sum obtained by applying $({\bf s}_{\alpha+})^{-1}$ and $({\bf s}_{\alpha-})^{-1}$, respectively.} \label{Paper2p4} 
\end{figure}

The intuitive difference between ${\bf s}_{\alpha +}$ and ${\bf s}_{\alpha -}$ is the following: given a major $\Phi$ with several singularities on the same ray in the direction $\alpha$, as on Fig. \ref{Paper2p4}. Take the ``left-most" singularity $\omega_0$ on this ray and draw a cut from it in the direction $\alpha$. We have a choice how to let this cut encircle the other singularities on the ray $\omega_0 \alpha$: either in the clockwise or in the counterclockwise direction. Depending on this choice, in general, different singularities will be visible on the first sheet. The first, resp., second choice gives the formal sum of microfunctions corresponding to $({\bf s}_{\alpha +})^{-1}\Phi$, resp., $({\bf s}_{\alpha -})^{-1}\Phi$.

Following ~\cite[p.188, R\'es I]{CNP}, but changing their notation, consider the map 
$$\bigsigma_{\alpha} := {\bf s}_{\alpha_+}\circ ({\bf s}_{\alpha_-})^{-1} \ : \ {\dot {\pmb{\cal R}}}(A) \ \to \ {\dot{\pmb{\cal R}}}(A)$$
called the {\bf ``homorphisme de passage''}. The map  $\bigsigma_\alpha$ is determined by its values on elementary resurgent sumbols, respects sums and convolution products of majors.


Denote, further, $\underline{S}_{\alpha} = \bigsigma_\alpha - 1$. 
The {\bf alien derivative} is defined as $\underline{\Delta}_\alpha = \Ln ({\bigsigma}_\alpha) = \sum_{n=1}^\infty \frac{(-1)^{n-1}}{n}(\underline{S}_{\alpha})^{n}$.
There is an obvious decomposition  
$$\underline{\Delta}_{\alpha} = \sum_{\omega \in \C} {\dot\Delta}_\omega \ , $$
where ${\dot \Delta}_\omega$ associates to a resurgent symbol $\dot\phi$ the microfunction centered at $\omega$  contained in the decomposition of $\underline{\Delta}_\alpha \dot\phi$ and the index $\alpha$ is suppressed on the right hand side. By shifting its argument, we can make this microfunction centered at zero, 
$$ ({\dot\Delta}_\omega \phi)(\xi) = (\Delta_\omega \phi)(\xi-\omega), $$
or, abusing notation, ${\dot\Delta}_\omega = e^{-\frac{\omega}{h}}\Delta_\omega$.

If $\phi^\omega$ is a microfunction centered at $\omega\in\C$, then the support of the formal resurgent symbol $\underline{\Delta}_{\alpha} \phi^\omega$ lies entirely on the ray emanating from $\omega$ in the direction $\alpha$.


Equations involving alien derivatives -- so-called ``alien differential equations" -- will be used in section \ref{Alienq1}.

\subsubsection{Mittag-Leffler sum} \label{MLS}

The concept of a Mittag-Leffler sum formalizes the idea of an infinite sum of resurgent functions $\sum_j \varphi_j(h)$ where $\varphi_j(h)$ have smaller and smaller exponential type, e.g., $\varphi = O(e^{-c_j/h})$ for $c_j\to \infty$ as $j\to \infty$.

Following ~\cite[Pr\'e I.4.1]{CNP}, let $\Phi_j$, $j=1,2,..$, be endlessly continuable holomorphic functions, $\Phi_j\in{\cal O}(\Omega_j)$, where $\Omega_j$ are sectorial neighborhoods of infinity satisfying $\Omega_j\subset \Omega_{j+1}$ and $\bigcup_j \Omega_j = \C$. Then ~\cite{CNP} show that there is a function $\Phi\in\Omega_1$ such that $$\Phi - \sum_{j=1}^n \Phi_j \ \in \ {\cal O}(\Omega_{n+1}), \ \ \ n=1,2,... . $$
In this case we will call $\Phi$ the {\bf Mittag-Leffler sum} of $\Phi_1,\Phi_2,...$.  

\subsection{Borel summation. Resurgent asymptotic expansions.} \label{BorelSum}

{\bf Definition.}  A {\bf resurgent hyperasymptotic expansion}  is a  formal sum 
$$\sum_k e^{-c_k/h}(a_{k,0}+a_{k,1}h+a_{k,2}h^2+...),$$ where: \\
i) $c_k$ form a discrete subset in $\C$ in the complement to some sectorial neighborhood of infinity in direction $\check A$;\\
ii) the power series of every summand satisfies the Gevrey condition, and \\
iii) each infinite sum $a_{k,0}+a_{k,1}h+a_{k,2}h^2+...$ defines, under formal Borel transform 
$$ {\cal B} \ : \ e^{-c_k/h }h^\ell \ \mapsto \ (\xi-c_k)^{\ell-1} \frac{\log (\xi-c_k)}{2\pi i \Gamma(\ell)} \ \ \text{if} \ \ell\in \N,$$ 
$$ {\cal B} \ : \ e^{-c_k/h } \ \mapsto \ \frac{1}{2\pi i(\xi-c_k)}, $$ 
an endlessly continuable microfunction centered at $c_k$. 

The authors of ~\cite{CNP} denote by  $\dot{{\cal R}} (A)$ (regular, as opposed to the bold-faced, ${\cal R}$) the algebra of resurgent hyperasymptotic expansions.

\begin{figure}[h]\includegraphics{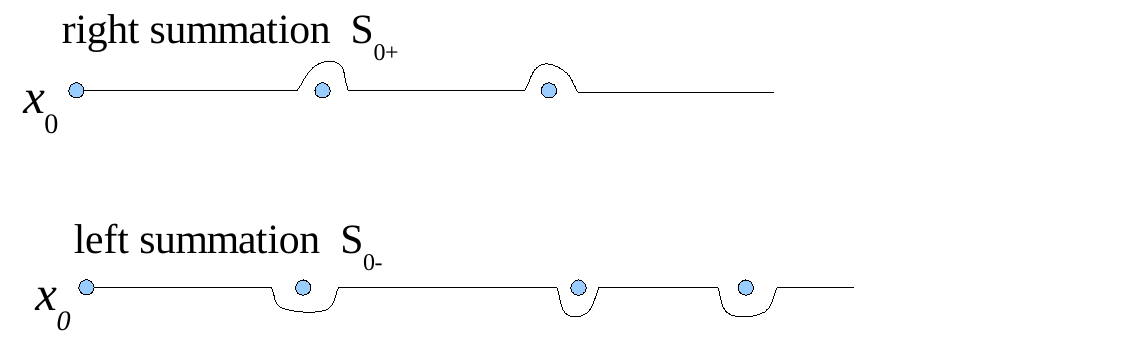} \caption{Integration paths for the left and right summation.} \label{nuthesisp3}
\end{figure}

The right and left summations of resurgent asymptotic expansions are defined in ~\cite{DP99} or ~\cite{CNP} as follows. Given a Gevrey power series $\sum_{k=1}^{\infty} a_k h^k$, replace it by a function (the corresponding {\it``minor''}) ${\pmb f}(\xi)=\sum_{k=1}^{\infty} a_k \frac{\xi^{k-1}}{(k-1)!}$, assume that ${\pmb f}(\xi)$ has only a discrete set of singularities, and consider the Laplace integrals $\int_{[0,\alpha)}e^{-\xi/h}{\pmb f}(\xi)d\xi$ along a ray from $0$ to infinity in the direction $\alpha$ deformed to avoid the singularities from the right or from the left, as on the figure \ref{nuthesisp3}.   

After some technical discussion, this procedure defines a resurgent function of $h$ which ~\cite{CNP} denote 
${\rm S}_{\alpha \pm} \left(\sum_{k=1}^{\infty} a_k h^k \right)$ and we, for typographical reasons, will denote ${\cal B}_{\alpha \pm} \left(\sum_{k=1}^{\infty} a_k h^k \right)$. Comparing the results of the left and right resummations lead to the notion of the ``homomorphisme de passage" discussed earlier.

\begin{Prop} If a resurgent microfunction is given by an asymptotic expansion in integer powers of $h$, then it defines a resurgent function for all $\arg h$. 
\end{Prop}

\begin{figure}[h]
\includegraphics{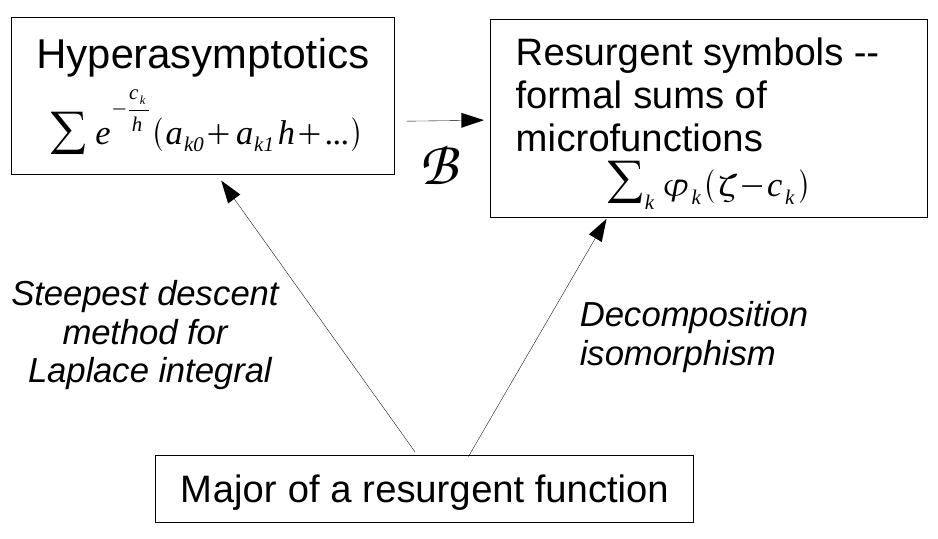} \caption{Logical relationship of different concepts in resurgent analysis.}
\end{figure}

\subsection{Small resurgent functions.} \label{SmallResFun}

{\bf Definition.} (~\cite[Pr\'e II.4, p.157]{CNP}) A microfunction $\varphi \in {\cal C}(A)$ is said to be a \underline{small microfunction } if it has a representative ${\bf\Phi}$ such that ${\bf\Phi} = o (\frac{1}{|\zeta|})$ uniformly in any sectorial neighborhood of direction $\check A$ for  $A' \subset\subset A$. 

For example, $h^\alpha$ for  $\alpha>0$  satisfies that property.

The following definition has been somewhat modified compared to (\cite[R\'es II.3.2, p.219]{CNP}).

\begin{figure}[h]
\includegraphics{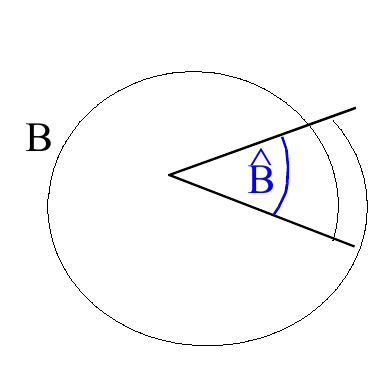} \caption{Arcs in the definition of a small resurgent microfunction.} \label{ExNote3p6}
\end{figure}

{\bf Definition.} For a given arc of direction $A$,  a \underline{small resurgent function} is such a resurgent function that all singularities of its major $\omega_\alpha$ satisfy $\re \omega_\alpha > 0$, except maybe for one $\omega_0=0$,  and if $\omega_0=0$ then the corresponding microfunction is small in the direction of a large (i.e. $>2\pi$) arc $B$ with $\hat B \supset A$, see figure \ref{ExNote3p6}. 

For example, $h^\alpha e^{-c/h}$ defines a small resurgent function for any $\alpha$ and $c>0$.

\begin{Lemma}  \label{MajorSmallResurgFctn} (\cite{G1}) A small resurgent function can be represented by a major that is $o(1/|\xi|)$ around the origin.
\end{Lemma}

We refer to \cite{G1} for a proof that if $g(t)$ is a holomorphic function in the neighborhood of the origin and $\varphi(h)$ is a small resurgent function, then $g(\varphi(h))$ is again resurgent. In particular, in this case the function $1/(c+\varphi(h))$ is resurgent for $0\ne c\in \C$, and therefore functions representable by resurgent asymptotic expansions ${\cal B}\left\{ e^{-c/h}(a_0+a_1 h + a_2h^2 +...) \right\}$ with $a_0\ne 0$ have a resurgent inverse. We do not know if an arbitrary nonzero resurgent function has a resurgent inverse.


\section{Resurgent solutions of a differential equation.} \label{ResurgDiffEqu}

\subsection{Existence problem} \label{ExistProb}

The book ~\cite{ShSt} discusses the resurgence properties of solutions of an equation
\begin{equation} h^m \frac{\partial^m \psi}{\partial q^m} + h^{m-1} P_{m-1}(q) \frac{\partial^{m-1} \psi}{\partial q^{m-1}} + ... + P_0(q) \psi = 0 \label{GenOrigEqu}, \end{equation}
where $P_j(q)$ are entire functions on $\C$; or, equivalently, one discusses why the Laplace transformed equation
\begin{equation} \left( \frac{\partial}{\partial \xi} \right)^{-m} \frac{\partial^m \Psi}{\partial q^m} + 
P_{m-1}(q) \left( \frac{\partial}{\partial \xi} \right)^{-m+1} \frac{\partial^{m-1} \Psi}{\partial q^{m-1} }  + ... + P_0(q)\Psi = 0 \label{eq8} \end{equation} 
has $m$ linearly independent multivalued endlessly continuable solutions $\Psi_j(q,\xi)$ giving resurgent solutions of \eqref{GenOrigEqu} via $\psi_j(q,h)={\cal L}\Psi_j(q,\xi)$.

The author has been unable to verify certain claims in the proof in ~\cite{ShSt} but believes that their result is correct. We are also aware of existence of the preprint ~\cite{Ec5}, addressing similar questions. The author hopes to offer a further exposition of this background material in his future work. 


The methods of ~\cite{ShSt} should equally well work to prove the existence result in the case when $P_0,...,P_{m-1}$ are analytic with respect to $q$ in the whole complex plane and depend polynomially on $h$. This covers the equations of the form
\begin{equation}  [-h^2 \partial^2_q + V_0(q) + hV_1(q)]\psi = E\psi,  \label{SchroExtraTerm} \end{equation}
where either $E$ is a complex number (``energy level''), or $E=hE_r$ and $E_r$ is a complex number (``rescaled energy''). In these cases we expect that the major of $\psi(E,q,h)$, resp., $\psi(E_r,q,h)$  can be chosen to holomorphically depend on $E$, respectively, on $E_r$,  and on $q$. 

We proved in ~\cite{G1} that this would imply that the equation  (\ref{SchroExtraTerm}) also has resurgent solutions when $E=h\varphi(h)$ and $\varphi(h)$ is a small resurgent function.


\subsection{Towards the notion of a parameter-dependent resurgent function.}

Solutions of equations of type \eqref{GenOrigEqu} should provide examples of parameter-dependent resurgent functions. An ultimate treatment of this notion should perhaps wait until all the details in ~\cite{ShSt}'s or other authors' approach to \eqref{eq8} have been clarified and we precisely know the properties of functions $\Psi(\xi,q)$, therefore we will restrict ourselves here to some preliminary remarks.

First of all, resurgent functions were defined in  section \ref{ResurFunDef} as equivalence classes of true functions modulo functions of sub-exponential decay for $h\to 0$. If we want to introduce a function $\psi(q,h)$ which is, say, analytic with respect to $q$ and resurgent with respect to $h$, we can either: 
\\ 1) require that for any fixed value of $q$  the function $\psi(q,h) \ \mod {\cal E}^{-\infty}$ is resurgent in $h$, that $\psi(q,h)$ analytically depends on $q$ as a true function of $h$, and not assume that majors $\Psi(q,\xi)$ have anything to do with each other for different values of $q$, or 
\\ 2) require that there exist a choice of majors $\Psi(q,\xi)$ (as true functions of $\xi$, not as classes modulo entire functions) which analytically depend on $q$ for $q$ in some open set in $\C$ and $\xi$ on a Riemann surface where all corresponding $\Psi(q,\xi)$ are simultaneously defined. \\
We prefer the latter concept, as it allows us to define a derivative of a resurgent function with respect to a parameter easily, and we hope that this property will eventually be established for solutions of (\ref{eq8}) as functions of $q$ and of the coefficients of the equation (\ref{eq8}). Working with the former concept is probably more difficult because differentiation with respect to $q$ does not a priori preserve the ${\cal E}^{-\infty}$-condition with respect to $h$.

The next issue is substitution of a resurgent function $\phi(h)$ into a parameter $E$ of a resurgent function $\psi(E,h)$ whose major $\Psi(E,\xi)$ analytically depends on $E$. The choice of $E$ in this notation should be suggestive of the main example -- when $E$ is a parameter in the Schr\"odinger equaton, say, (\ref{WL}); we will for the moment suppress the dependence on $q$. Substitution of $E=\phi(h)=E_r h$ where $E_r$ is a complex number is routinely done in ~\cite{DDP97} and ~\cite{DP99}, and we will not discuss it here. It is important, however, that locations of singularities of majors $\Psi_r(E_r,\xi)$ of $\psi_r(E_r,h)=\psi(E_rh,h)$ do not depend on $E_r$, and it is expected from the general theory that majors $\Psi_r(E_r,\xi)$ are analytic with respect to $E_r$. In ~\cite{G1} we have proven that for a small resurgent function $\phi(h)$ the composite function $\psi_r(\phi(h),h)$ is again resurgent and has a hyperasymptotic expansion that one would  expect from formal manipulation with hyperasymptotic expansions of $\phi$ and $\psi_r$.


\subsection{Formal resurgent symbol solutions of the Schr\"odinger equation.}  
\label{FormalSolusDefn}

Given now an $h$-differential operator $P(q,h\partial_q,h)$, one can consider ``actual'' 
\footnote{The word ``actual" is put here in the quotation marks because resurgent solutions are still formal objects defined $\mod {\cal E}^{-\infty}$, and they solve the differential equation hyperasymptotically, i.e. $\mod {\cal E}^{-\infty}$} 
 solutions of $P\varphi = 0$ that are resurgent with respect to $q$. On the other hand, we have formal solutions of this equation:

{\bf Definition.} $\phi(q) = e^{S(q)/h}(a_0(q) + h a_1(q) + .. )$ is called an {\it elementary formal resurgent (WKB) solution} of $P\phi$ if $\phi(q)$ defines a resurgent microfunction for any $q$ and if $a_0(q) + ha_1(q) + ...$ satisfies the equation $e^{S(q)/h}Pe^{-S(q)/h}$ in the sense of formal power series in $h$, and the Borel sum of the series defines a resurgent microfunction for every $q$ in an appropriate domain; \\
A {\it formal resurgent symbol solution}, or, for brevity, {\it formal solution}, is a formal sum of  elementary formal solutions which defines, via decomposition theorem, a resurgent function for every value of $q$ in the considered domain.

Now let us specialize to a Schr\"odinger-type equation 
\begin{equation}  P\psi(q,h) \ :=  - h^2 \partial^2_q \psi(q,h) + V(q,h)\psi(q,h) \ = \ 0, \label{STE} \end{equation} 
where $V(q,h)=V_0(q) + hV_1(q)$. Define a {\bf turning point} of $P$ as a $q_*$ such that $V_0(q_*)=0$. A turning point is called simple, double, etc., depending on the multiplicity of this zero.

We will define the {\bf classical momentum} by $p(q)=\sqrt{-V_0(q)}$; two determinations of this square root define a two-sheeted ramified cover of the complex plane of the variable $q$, which is customarily referred to as the {\bf Riemann surface of the classical momentum}. Fixing a point $q_0$ on the Riemann surface of $p(q)$, we further consider a {\bf classical action} $S(q,h)=\int_{q_0}^q p(q') dq'$ which is defined on the universal cover of $\C$ with the turning points removed. 

In a neighborhood of every $q\in \C$ that is not a turning point, there are two linearly independent formal solutions of (\ref{STE}) of the form 
$$ \phi(q,h) \ = \ e^{\frac{iS(q)}{q}}(a_0(q)+a_1(q)h+a_2(q)h^2+... ). $$
We will say that $\phi$ corresponds to the first or to the second sheet of the Riemann surface of $p(q)$ depending in whether $dS(q)/dq$ coincides with first or the second sheet value of $p(q)$. 

Once the result about existence of resurgent solutions for (\ref{STE}) is established, resurgence of formal solutions $\phi(q,h)$ will follow as well.

The correspondence between actual and formal solutions of a differential equations is not straightforward is the subject of the next subsection.

\subsection{Stokes Phenomenon.}

\begin{figure}
\includegraphics{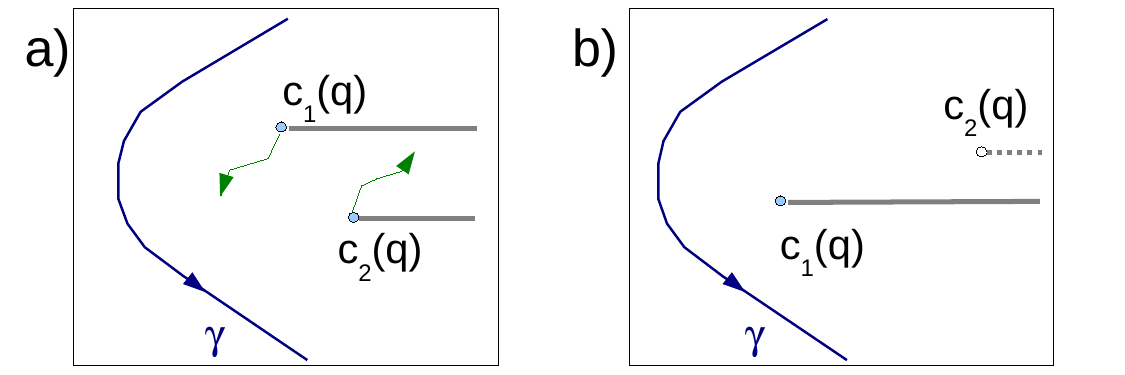} \caption{On this figure the first sheet of the Riemann surface of $\Psi(q,\xi)$ is shown for fixed $q$ in projection to the complex plane of $\xi$. Suppose $\Psi(q,\xi)$ has two singularities $c_1(q)$ and $c_2(q)$ present on the first sheet (part a) and therefore $\psi(q,h)$ has a hyperasymptotic expansion of the form $e^{-c_1/h}(...)+e^{-c_2/h}(...)$. As the value of $q$ changes, the singularities move along the shown trajectories, and for a different value of $q$ (part b) the singularity $c_2(q)$ has moved to under the cut starting at $c_1$ to the second sheet; in this situation $\psi(q,h)\sim e^{-c_1(q)/h}(...)$. } \label{StPhenFig}
\end{figure}

Assume $\psi(q,h)=\int_\gamma \Psi(q,\xi) e^{-\xi/h} d\xi$ be a Laplace integral representing a $q$-dependent resurgent function $\psi$ and $\Psi$ and $\psi$ both depend analytically on $q$ in a suitable sense. Then we expect that locations of singularities of $\Psi$ move continuously with respect to $q$, which may result in that the collection of singularities visible on the first sheet will start changing and this result in an apparent discontinuity of the hyperasymptotic expansion with respect to $h$; see example on figure \ref{StPhenFig}. 

\begin{figure}
\includegraphics{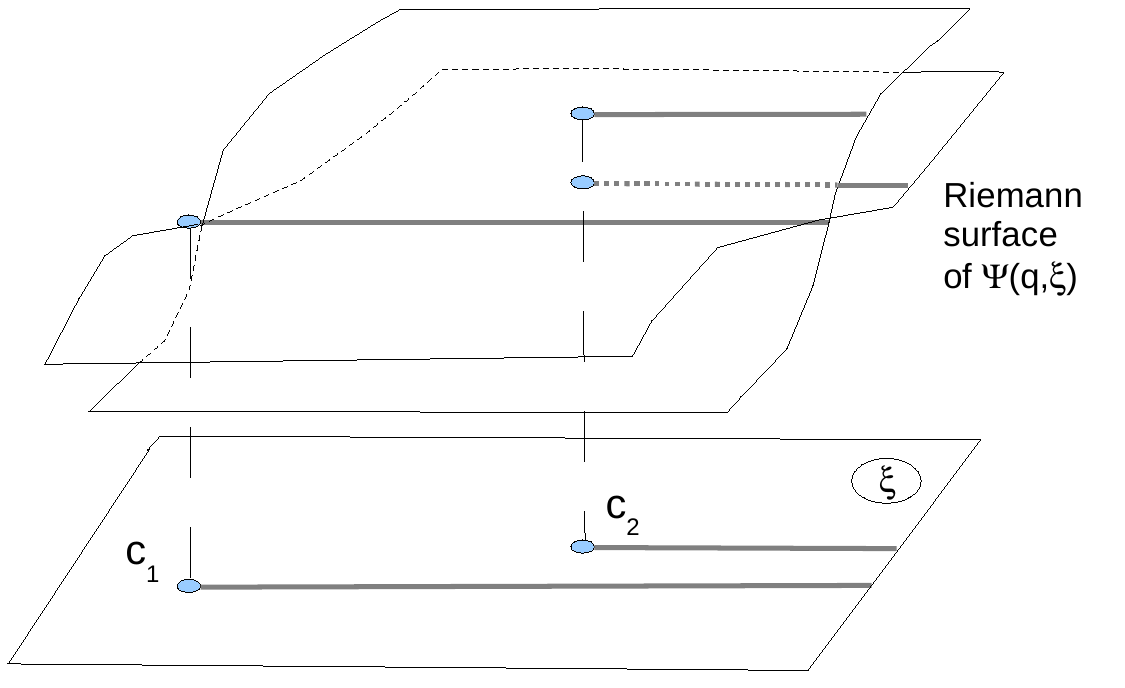} \caption{Two equal singularities at $c_2(q)$: one appearing on and the other disappearing from the first sheet; $c_1$ is drawn as independent of $q$ for simplicity.} \label{DecoupledSing}
\end{figure}

Not every time, however, when $c_2(q)$ crosses the cut starting at $c_1(q)$ the hyperasymptotic expansion has to change. It is also possible that the singularity that disappears from the first sheet equals the singularity that simultaneously appears on the first sheet, figure \ref{DecoupledSing}.   In this situation ~\cite{V83} says that the singularities at $c_1(q)$ and $c_2(q)$ are {\it decoupled} and that there is no Stokes phenomenon.

According to ~\cite{V83}, for an equation of type (\ref{STE}) the Stokes phenomenon can only happen when $q$  crosses one of at most countably many curves on the complex plane of $q$. More specifically, let us give the following

{\bf Definition.} A {\bf Stokes curve} corresponding to $P$ and to resummation direction $\alpha$ is a curve on the Riemann surface of the classical momentum given by $\arg [iS(q)-iS(q_0)] =\alpha$, where $S(q)$ is the classical action.

Stokes curves either go from a turning point to infinity, or end in an other turning point. In the last case we call them {\bf double}, or {\bf bounded Stokes curves}. Stokes curves split the complex plane of $q$ into {\bf Stokes zones} or {\bf Stokes regions}.

The {\bf canonical length} of an unbounded Stokes curve is defined to be infinite, and that of a bounded Stokes curve is taken to be the absolute value of $\int p(q')dq'$ along this curve.

Fix a Stokes zone $A$. It is explained in ~\cite{V83} that if $\psi_+(q,h)$, $\psi_-(q,h)$ are elementary formal resurgent solutions of (\ref{STE}) corresponding to different determinations of the classical momentum and if $\psi(q,h)$ is an ``actual'' resurgent solution of (\ref{STE}), then there are resurgent symbols $A_+$, $A_-$ (which we will by abuse of notation write as if they were functions of $h$) such that a hyperasymptotic expansion
$$ \psi(q,h) \ \sim \ A_+(h) \psi_+(q,h) + A_-(h)\psi_-(q,h) $$
holds for all $q$ inside $A$. In the resurgent literature this idea is usually expressed by saying that {\it away from the Stokes curves analytic continuation of formal solutions of a differential equation corresponds to analytic continuation of actual 
solutions.}

Given two Stokes zones $A$ and $B$ and two hyperasymptotic expansions of the same ``actual" resurgent solution $\psi(q,h) \ \sim \ A_+(h) \psi_+(q,h) + A_-(h)\psi_-(q,h)$ and $ \psi(q,h) \ \sim \ B_+(h) \psi_+(q,h) + B_-(h)\psi_-(q,h) $ valid in $A$ and $B$ respectively, the {\bf connection problem} consists in finding a $2\times 2$ matrix of resurgent symbols such that 
$$\left( \begin{array}{c} B_+ \\ B_- \end{array} \right) \ = \ \left( \begin{array}{cc} c_{++} & c_{+-} \\ c_{-+} & c_{--} \end{array} \right)  \left( \begin{array}{c} A_+ \\ A_- \end{array} \right). $$
This matrix is called a {\bf connection matrix}.

\begin{figure}[h]\includegraphics{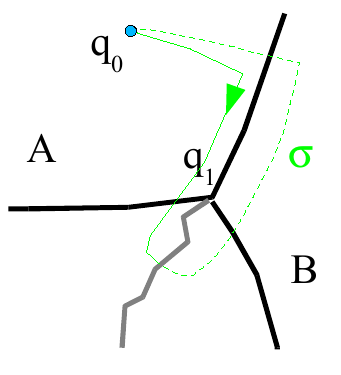} \caption{The path used in the calculation of the connection coefficient in Theorem \ref{ThmVoros}. As $q_1$ is a branch point of the classical momentum $p(q)$, we drew a corresponding branch cut.  } \label{RDRWp19}
\end{figure}

Typically one solves the connection problem for neighboring Stokes zones first. The basic result is the following:

\begin{Thm} \label{ThmVoros} (see ~\cite{V83}) Suppose $q_1$ is a simple turning point, and $L$ is an unbounded Stokes curve emanating from $q_1$. Consider two elementary formal WKB solutions $\psi_-$, $\psi_+$ corresponding to two opposite determinations of classical momentum and normalized to be $1$ at a point $q_0$ in the Stokes region $A$. Let $\psi_+$ be exponentially growing, and $\psi_-$ exponentially decreasing away from $q_1$ along $L$. Then a solution representable by a formal resurgent symbol $A_+ \psi_+ +A_-\psi_-$ in the Stokes region $A$ will be representable by a formal resurgent symbol $A_+\psi_+ + (A_-+CA_+)\psi_-$ in the Stokes region $B$, where $C$ is the monodromy of the formal solution $\psi_+$ along the path $\sigma$, see figure \ref{RDRWp19}. \end{Thm}

The equation (\ref{WL}) that will occupy us in this article has double turning points, and the derivation of connection formulae for that case from Theorem \ref{ThmVoros} is the subject of section \ref{ConnectionFlae}.

If a differential equation (\ref{STE}) admits two linearly independent solutions representable (away from the turning points) by resurgent hyperasymptotic power series expansions with respect to $h$, then its connection coefficients between different Stokes regions will be automatically resurgent. Indeed, calculation of these connection coefficients involves splitting a major of a resurgent solution $\psi(q,h)$ into resurgent microfunctions and dividing those microfunctions by resurgent microfunctions of the type ${\cal B}(e^{S(q)/h}(a_0(q)+a_1(q)h+...)$ representing formal WKB solutions of (\ref{STE}) and having resurgent inverses by section \ref{SmallResFun}.


\section{Plan of the proof of the main result.} \label{PlanOfProof}

In order to solve the eigenvalue problem for the equation
\begin{equation} \label{WL} [- h^2 \frac{d^2}{dq^2} + (f')^2 - h f'']\psi(q,h) = E\psi(q,h), \ \ \ \ E=O(e^{-const/h}),   \end{equation}
and to prove theorem \ref{MainResult}, we start by studying in section \ref{FormalWKBSolus} its formal solutions. Unlike actual solutions of (\ref{WL}), its formal solutions are not in general univalued functions of $q$, and they can have monodromies around various loops in the $q$-plane, discussed in subsection \ref{MonodrExp}.
 
In order to pass from formal to actual solutions of (\ref{WL}), we need to study the geometry of its Stokes curves in section \ref{StPatternConnProb}. For $E=hE_r=O(e^{-const/h})$, the turning points of (\ref{WL}) will be double, and the derivation of relevant connection formulas will be given in section \ref{ConnectionFlae}.

In section \ref{TransMtx} we will define a transfer matrix $F$ dependent on $E_r$ which is basically the product of connection matrices across the double turning points. It is in terms of $F$ that we will write the quantization condition (\ref{qCond81}) -- the condition on $E_r$ that (\ref{WL}) has an actual solution satisfying $\psi(q)=\psi(q+1)$. In section \ref{ConnMatQuantCond} we will rewrite the quantization condition as a polynomial in terms of quantities $\mu_j$ and $\tau_j$ that are, morally, $const\cdot E_r$ and $const\cdot E_r e^{-\frac{const}{h}}$.

The section \ref{ResurgTranscEqu} deals with the question how to solve an equation of such form for $E_r$. We will draw a Newton polygon of the equation by plotting a term $E^j e^{k/h}$ as a point with coordinates $(j,k)$ and present an iterative procedure  to recursively obtain smaller and smaller $e^{-const/h}$-order terms in the hyperasymptotic expansion of $E_r$. The outcome of the section \ref{ResurgTranscEqu} is that the quantization condition that we set up treating $E_r$ as a complex number, can be solved for $E_r$ and the solution will be a resurgent function in $h$ if all the ingredients $\mu_j$ and $\tau_j$ of the quantization condition are resurgent (plus some additional technical conditions). Note that we can substitute small resurgent functions for $E_r$ in the equation (\ref{WL}) for $E=hE_r$, as we discussed in detail in ~\cite{G1}.

We finish the proof by plotting in section \ref{Solving} the Newton polygon corresponding to the quantization conditon (\ref{QCond}) and concluding that this quantization condition has a correct number of exponentially small solutions.


\section{Formal WKB solutions and formal monodromies} \label{FormalWKBSolus}

From now on we let the function $f(q)$ be a polynomial in $\sin 2\pi q$ and $\cos 2\pi q$, real for real $q$,  we let $f$ have  $n$ real local minima $q_1,...,q_{2n-1}$ and $n$ real local maxima $q_2,..,q_{2n}$ on the period, where $0<q_1<q_2<...<q_{2n-1}<q_{2n}<1$. We require $f''(q_j)\ne 0$. 

A monodromy of an elementary formal WKB solution along some path $\rho(t)$ on a universal cover $\tilde\C$ of $\C\backslash\{\text{turning pts}\}$  is defined as $\psi(\rho(1))/\psi(\rho(0))$ and will be denoted $\exp [2\pi i s_\rho]$. Since this expression is a quotient of two resurgent microfunctions representable in the form $e^{S(q)/h}(const+O(h))$, it is a resurgent microfunction and therefore the power series in $h$ representing $s_\rho$ is resurgent.

We are going to calculate various monodromy exponents  $s_\gamma$, $s_\delta$, etc. as a resurgent symbol, and therefore we do not therefore care about Stokes phenomena in this section.

\subsection{Cuts, signs and branches.}

In this section we will discuss  formal WKB solutions of 
\begin{equation} P\psi \ := \ \left[ -h^2 \partial^2_q + (f')^2 - hf''\right]\psi \ = \ E \psi, \label{WLcsb} \end{equation}
where $E$ is a complex number.

For $E\ne 0$ and $|E|$ sufficiently small, the classical momentum $p(q)=\sqrt{E-(f'(q))^2}$ is defined on a two sheeted cover of the complex plane of $q$. For $E=0$, the two determinations of $p(q)$ are $\pm f'(q)$, and one can think of the Riemann surface of $p(q)$ as of two separate sheets having contact at points $q_j$ where $f'(q_j)=0$.

The formulas related to formal solutions of the equation \ref{WLcsb} can be established, for definiteness, for $E>0$, and then analytically continued to other values of $E$, whenever appropriate.

\begin{figure}[h]\includegraphics{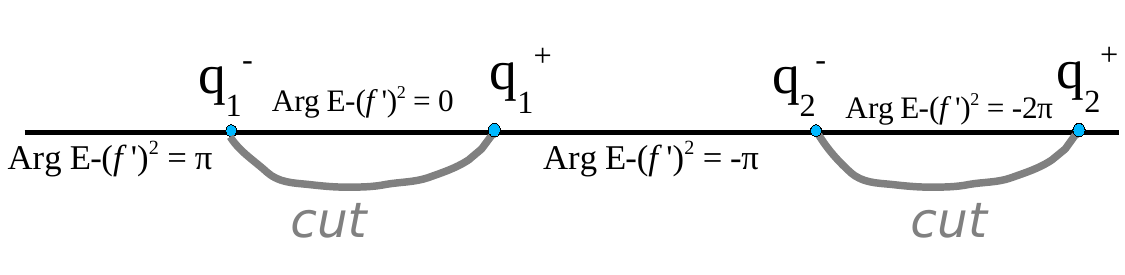} \caption{Choice of $\Arg E-(f')^2$} \label{nuthesisp5}
\end{figure}

When $E>0$ and $|E|$ is sufficiently small, the double turning points $q_j$ on the real axis for $E=0$ split into pairs $q_j^-<q_j<q_j^+(<q_{j+1}^{-})$ of simple turning points still on the real axis. The Riemann surface of $\sqrt{E-(f')^2}$ will be described as the plane with cut connecting $q_j^-$ to $q_j^+$ and going a little below the real axis. To specify the determination of $p(q,E)$ on the first sheet, we define $\Arg (E-(f')^2)$ for real values of $q$ on figure \ref{nuthesisp5}. As $E\to 0$, on the first sheet $i p(q,E)\to f'(q)$.

\subsection{Formal solutions.}

In order to find a formal WKB solution of (\ref{WLcsb}), we will be looking for a series 
$$y(h,q) = y_0(q) + hy_1(q) + h^2 y_2(q) + ... $$
solving 
$$ (P-E)\left\{ \exp \left\{ \int^q \frac{i}{h}\sqrt{E-(f')^2} + y(q')dq'\right\} \right\} = 0 .$$

We arrive at the Riccati equation
$$  -2y(q) i\sqrt{E-(f')^2}  - h y(q)^2 +  
i\frac{f'f''}{\sqrt{E-(f')^2}} - h y'(q)   - f'' = 0 , $$ 
that allows to calculate $y_j$'s recursively:
$$  y_0(q) =  \frac{f'f''}{2(E-(f')^2)}    - \frac{f'' }{2i\sqrt{E-(f')^2}}, $$ 
etc.

There is a following way to repackage this series in $h$. Let us argue more generally for an equation
\begin{equation} -h^2 \partial^2_q \psi + V(q)\psi = E\psi  \label{eq10} \end{equation}
where $E\in \C$ and $V(q)=V_0(q)+hV_1(q)$.

\begin{Prop} \label{secondWay} There is a power series ${\tilde y}(q)= {\tilde y}_0(q)+h{\tilde y}_1(q)+h^2{\tilde y}_2(q)+...$  such that
\begin{equation} \psi = \frac{1}{\sqrt{ \sqrt{E-V(q)} + h{\tilde y}}} \exp \left( \frac{i}{h} \int^q \left\{ \sqrt{E-V(q)} + h{\tilde y} \right\} dq \right)  \label{tildey} \end{equation}
is a formal solution of (\ref{eq10}).  Moreover:\\
1) Changing the determination of $\sqrt{E-V(q)}$ changes the sign of $\tilde y$; \\
2) $\tilde y$ contains only odd  powers of $h$, starting from $h^1$, i.e. ${\tilde y}_{2k}=0$.  
\end{Prop}

\textsc{Proof.}  Solving a Riccati equation corresponding to (\ref{eq10}), e.g. ~\cite[\S 2.1]{KT}, we can obtain a power series $y(q)=y_0 + y_1 h + ...$ such that 
$$ \exp \left[ \int dq \left\{ \frac{i}{h} \sqrt{E-V(q)} + y(q) \right\} \right] $$
is a formal WKB solution of (\ref{eq10}). Knowing $y(q)$ and the relation 
$$   y(q) \ = \  \frac{i}{h}\left\{\sqrt{E-V(q)}-\sqrt{E-V_0(q)} + h{\tilde y}\right\} - \frac{1}{2} d\ln \left\{\sqrt{E-V(q)} + h{\tilde y}\right\}, $$
we can obtain a recursive relation for ${\tilde y}_j$'s. 

To prove the first property, substitute (\ref{tildey}) into (\ref{eq10}) and obtain after simplification: 
\footnotesize
$$ -h^2 \partial_q  \frac{(-1)}{2} \frac{\partial_q \left( \sqrt{E-V(q)} + h{\tilde y}\right)}{\left( \sqrt{E-V(q)} + h{\tilde y}\right)^{3/2}} 
+ \frac{\left(\sqrt{E-V(q)} + h{\tilde y} \right)^2}{\sqrt{\sqrt{E-V(q)}+ h{\tilde y}}}   
+ \frac{V(q)-E}{\sqrt{\sqrt{E-V(q)} + h{\tilde y}}} = 0. $$ \normalsize
It is obvious that simultaneous change of the determination of $\sqrt{E-V(q)}$ and of the sign of $\tilde y$ preserves this equality. 

This equality will also be preseved if we simultaneously change of sign of $h$ and the sign of $\tilde y$, therefore  $\tilde y$ contains only odd powers of $h$.  $\Box$.


\subsection{Calculation of monodromy exponents.} \label{MonodrExp}

Given a formal WKB solution $\psi(q,h)$ of our differential equation and a path $\rho(t)$, $0\le t \le 1$, on the Riemann surface of the classical momentum, we will define the {\bf formal monodromy} of $\psi$ along the path $\rho$ as $e^{2\pi i s_\rho} = \psi(\rho(1),h)/\psi(\rho(0),h)$ and call $s_\rho$ the {\bf monodromy exponent}. In case $\psi(q,h)$ is representable by a product $e^{S(q)/h}$ times a resurgent power series in $h$, the formal monodromy and the monodromy exponent will be resurgent as well. One can think of $s_\rho$ and $e^{2\pi i s_\rho}$ as of resurgent asymptotic expansions in $h$ or as of corresponding microfunctions; we will be suppressing this distinction.

\begin{figure}[h]
\begin{tabular}{ccc} 
\raisebox{-1cm}{\includegraphics{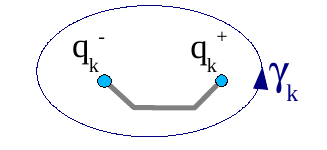}} &   \raisebox{-1cm}{\includegraphics{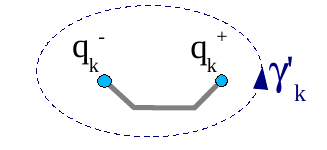}} \\
 \parbox{6cm}{$s_{\gamma_k}\approx -\frac{E}{2f''(q_k)h}-1 $  }  &    \parbox{6cm}{$s_{\gamma'_k}=-1-s_{\gamma_k}\approx \frac{E}{2f''(q_k)h}$ } \\
\raisebox{-1cm}{\includegraphics{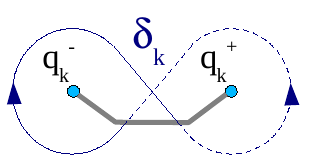}} &   \raisebox{-1cm}{\includegraphics{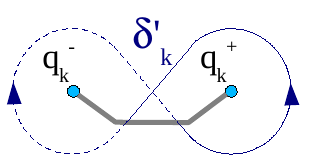}} \\
 \parbox{6cm}{$ s_{\delta_k}=-\frac{1}{2}-s_{\gamma_k} \approx \frac{E}{2f''(q_k)h} + \frac{1}{2}$  }  &    \parbox{6cm}{$s_{\delta'_k} =\frac{1}{2}+s_{\gamma_k}\approx -\frac{E}{2f''(q_k)h}-\frac{1}{2} $ }
\end{tabular} 
\caption{Various formal monodromy exponents} \label{MonodromyExp}
\end{figure}

As before, let $\gamma_k$ (resp., $\gamma'_k$) be a counterclockwise loop around the pair of turning points $q^+_k, q^-_k$ on the first (resp., second) sheet of the Riemann surface of the momentum, and denote $e^{2\pi i s_{\gamma_1}}$ and $e^{2\pi i s_{\gamma'_1}}$ the corresponding monodromies of formal WKB solutions along these loops.
Analogously, let $e^{2\pi i s_{\delta_k}}$ and $e^{2\pi i s_{\delta'_k}}$ be the monodromies of formal solutions along the figure-eight loops shown on the Fugure \ref{MonodromyExp}.

It is elementary to calculate directly the leading powers of $h$ in the power series representing resurgent symbols $s_{\gamma_k}$. In order to derive the corresponging expansions of  $s_{\gamma'_k}$, $s_{\delta_k}$, $s_{\delta'_k}$, one argues as follows for $\gamma'_k$, and similarly for $\delta_k$ and $\delta'_k$.

\begin{Lemma} \label{SumIsMinus1} 
We have  $s_{\gamma_k}+s_{\gamma'_k}=-1$. \end{Lemma}

\textsc{Proof.} Denote $V(q)=[f'(q)]^2-hf''(q)$ and use proposition \ref{secondWay}.  If we multiply  two formal solutions corresponding to different sheets, we will get $-\frac{1}{\sqrt{E-V(q)}+h\tilde y}$. Since $h\tilde y$ can be disregarded for the sake of this argument,
and $E-V(q)$ has two zeros inside the contour, the argument of this fraction turns by $-2\pi$ as we run around $\gamma_k$. Hence the lemma. $\Box$ 
 
Summarizing,

\begin{Prop} The monodromy exponents satisfy the formulas listed on the figure \ref{MonodromyExp}, where the sign $\approx$ means $\mod O(E^2/h) + O(h)$. \end{Prop}

\begin{figure}[h]\includegraphics{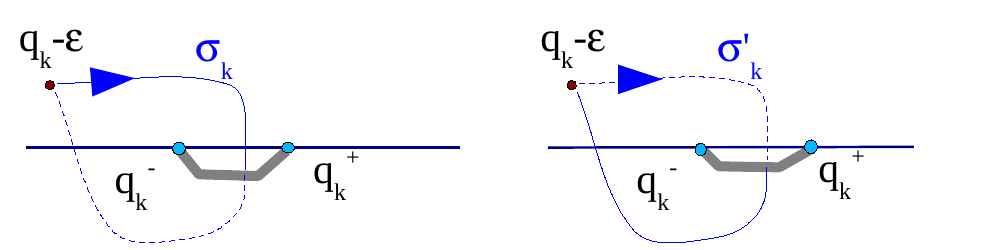}  \caption{Paths $\sigma_k$ and $\sigma'_k$.} \label{nuthesisp8}
\end{figure}

The knowledge about two more formal monodromies  will be needed in the later sections. Let $q_k-\varepsilon$ be close enough to $q_k$. 
The path $\sigma_k$ is defined for $E>0$ as starting at $q_k-\varepsilon$ on the first sheet, going under the cut between $q_k^-$ and $q_k^+$, and ending at $q_k-\varepsilon$ on the second sheet; the path $\sigma'_k$ is obtained from $\sigma_k$ by interchanging the sheets, figure \ref{nuthesisp8}.  

We easily obtain:

\begin{Lemma}  $e^{-2\pi i s_{\sigma_k}} = -e^{2\pi i s_{\sigma'_k}}$.
\end{Lemma}

Yet another elementary calculation shows that:

\begin{Prop} We have for $k$ odd
$$ \int_{\sigma_k} \left[  - \frac{f''}{2i\sqrt{E-(f')^2}} \right] dq \ =  \Ln \left[-\frac{2f'(q_k-\varepsilon)}{\sqrt{E}}\right](1+O(E)),$$
and for $k$ even 
$$ \int_{\sigma'_k} \left[  - \frac{f''}{2i\sqrt{E-(f')^2}} \right] dq \ =  - \Ln \left[\frac{2f'(q_k-\varepsilon)}{\sqrt{E}}\right](1+O(E)).$$
Here the branch of the logarithm is real for $q_k-\varepsilon$ on the real line to the left of $q_k^{-}$.
\end{Prop}

\begin{figure} \includegraphics{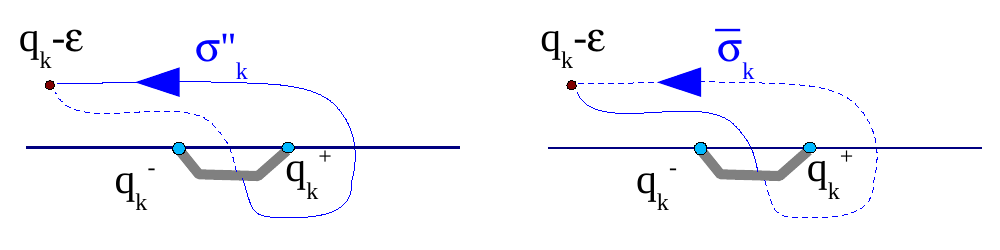} \caption{Paths $\sigma''_k$ and ${\bar\sigma}_k$.} 
\label{Paper2p12}  
\end{figure}

Finally, we will define paths $\sigma''_k$ and ${\bar \sigma}_k$ as on figure \ref{Paper2p12}.

By deforming integration contours one immediately sees that
\begin{equation} s_{\sigma''_k} = -s_{\sigma_k} + s_{\delta'_k}; \ \ \ \ s_{\bar \sigma_k}=-s_{\sigma'_k} + s_{\delta_k}. \label{SigmaBis} \end{equation}


\section{Stokes pattern and the connection problem.} \label{StPatternConnProb}

We are interested in the spectrum of the Witten Laplacian
$$ P \ = \ -h^2 \partial^2_q + (f')^2 - hf'' $$
with periodic $f$ and periodic boundary conditions on the eigenfunctions.  According to the usual philosophy (see, e.g., ~\cite{DDP97}), we will take the formal WKB solutions of this equation, consider the Stokes curves and the Stokes regions and solve the connection problems between different Stokes regions.

\begin{figure} 
\includegraphics{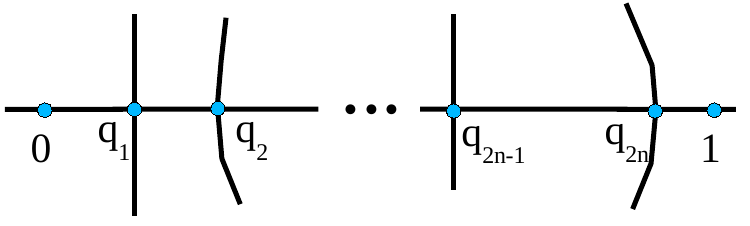} \caption{The Stokes pattern for the equation (\ref{WLspcp}) in case $\arg h=0$.} \label{Paper2p8}
\end{figure}

If $\arg h=0$, the Stokes pattern (i.e. the picture formed by all Stokes curves) for the equation
\begin{equation} (-h^2 \partial^2_q + (f')^2 - hf'' )\psi(q,h) \ = \ hE_r \psi(q,h) \label{WLspcp} \end{equation}
 is as shown on figure \ref{Paper2p8}. Namely, from every real turning point there will emanate four perpendicular Stokes curves two of which will go along the real line and connect nearby real turning points.

As is made clear in ~\cite{DP99}, it is easier to solve the connection problems when all Stokes curves are simple. Since there are only finitely many turning points inside a period $0\le \Re q < 1$, there is a number $\alpha_0>0$ that for $0<\arg h<\alpha_0$ there are no double Stokes curves.  Then, by continuity considerations customary in resurgent analysis, for these values of $\arg h$ the decomposition of solutions into microfunctions is the same as for $\arg h = 0+$. 

\begin{figure}[h]\includegraphics{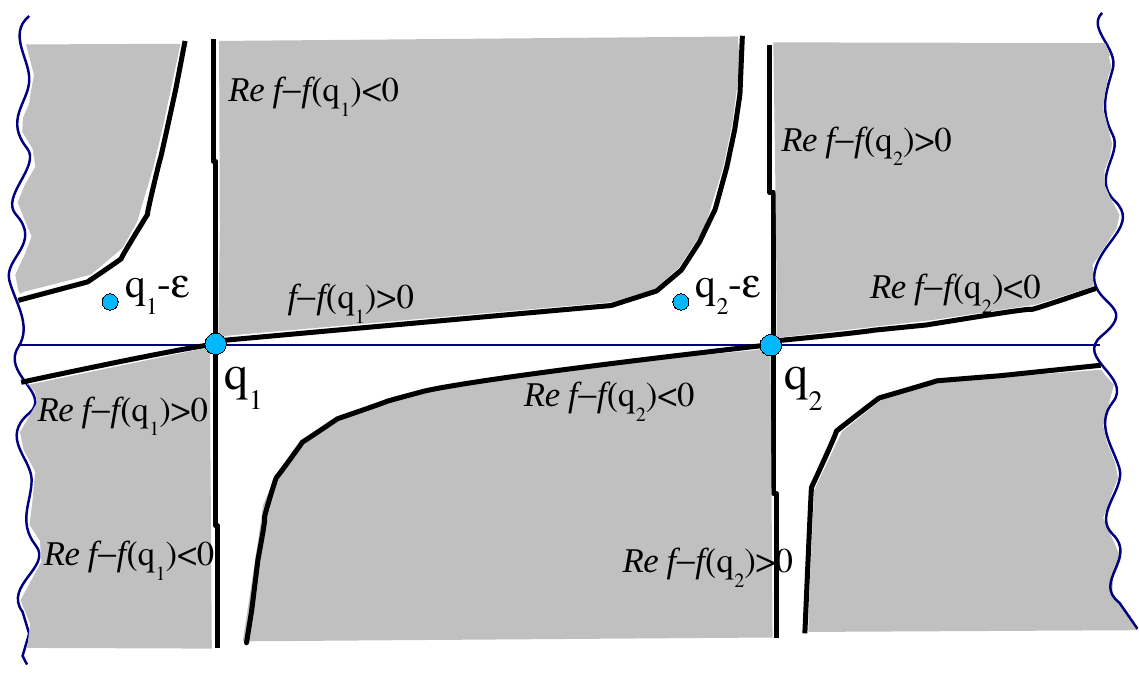} \caption{The Stokes pattern for $0<\arg h<<1$.} \label{nuthesisp10}
\end{figure}

Therefore from now on we shall fix such $\arg h>0$; the Stokes pattern will deform as shown on figure \ref{nuthesisp10}. We will further choose $\varepsilon$ so that $q_j- \varepsilon$ lie between Stokes curves. These points will be used in the calculation of the connection matrices across the double turning points. 
 

\subsection{Properties of the Stokes pattern}

The fact that for a polynomial $f$ the Stokes pattern does not have any pathologies, is almost obvious and well known. Our $f$ is not a polynomial, but instead a trigonometric polynomial $f(q) = P(\sin 2\pi q, \cos 2\pi q)$. Still, this is enough to have a nice Stokes pattern (without, e.g., dense families of Stokes curves), as we will now demonstrate. The condition that $f$ is a trigonometric polynomial in this paper is imposed in order to be able to prove these lemmas.

\begin{Lemma} If $f(q) = P(\sin 2\pi q, \cos 2\pi q)$, $P$ a polynomial, then for $0\le \Re q < 1$ there are only finitely many critical points.
\end{Lemma}

\textsc{Proof.} Indeed, by passing to $t=\sin 2\pi q$, one can find an algebraic equation satisfied by sines of all critical points of $f$. $\Box$

\begin{Lemma} Under above assumptions, all the Stokes curves are contained in the union of finitely many real curves given by algebraic equations on $\Re$ and $\Im$ of $\sin 2\pi q$. \end{Lemma}

\textsc{Proof.} The condition for a point $q$ to lie on a Stokes curve emanating from a turning point $q_j$ (real or not), $\arg [f(q) - f(q_j)] = \alpha$ can be translated into $\Re [f(q) - f(q_j)] = k  \Im [f(q) - f(q_j)]$, which in turn implies an algebraic equation on $\Re$ and $\Im$ of $\sin 2\pi q$. Since there are finitely many critical points $q_j$, and every real algebraic curve has finitely many connected components (~\cite[Th.3]{Wh}), the lemma follows. $\Box$.

\begin{Lemma} Under above assumptions there is $0<\theta_0 \ll 1$ such that for $0<|\arg h| < \theta_0$ there are no double Stokes curves. \end{Lemma}

\textsc{Proof} follows from the finiteness of the set $$\{ \arg [f(q_i) - f(q_j)] \ | \ q_i,q_j \ \text{turning pts} \}.$$ ~ $\Box$

%


\section{Connection formulae for a double turning point.} \label{ConnectionFlae}

Consider the equation 
\begin{equation} [-h^2\partial^2_q + (f')^2 -hf'']\psi \ = \ E\psi. \label{DeformedSchroe} \end{equation}
For $E\in \C$ and $0<|E|<<1$, the equation (\ref{DeformedSchroe}) pairs of simple turning points $q_j^-$ and $q_j^+$, dependent on $E$, which coalesce to double turning points $q_j$ when $E$ becomes zero or $hE_r$ for $E_r\in\C$ or $E_r$ a small resurgent function of $h$. We are going to imitate 
 ~\cite[section 4.1]{DP99}'s  method of passing to the limit $E\to 0$ and deriving connection formulae 
across Stokes curves emanating from $q_j$ for $E=hE_r$ based on connection formulas across Stokes curves emanating from $q_j^+$, $q_j^-$ for $E\ne 0$. We will see that ~\cite{DP99}'s argument does not change significantly, but a detailed discussion of ~\cite{DP99}'s method will stay outside of the scope of this article.

To simplify notation, we will argue in this section for double turning points $q_1$ where $f''(q_1)>0$ and $q_2$ where $f''(q_2)<0$; similar connection formulae hold true for other double turning points $q_j$ as well.

\begin{figure}[h]\includegraphics{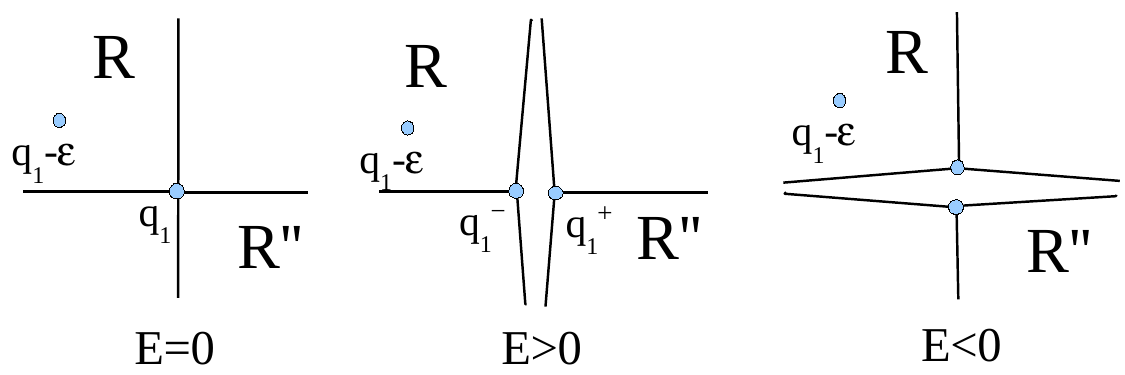} \caption{Deformation of a double turning point into a pair of simple turning points} \label{RDRWp18}
\end{figure}

Take a point $q_1$ for which $f'(q_1)=0$.  Consider the basis $\psi_-(q,E,h)$, $\psi_+(q,E,h)$ of formal resurgent WKB solutions of (\ref{DeformedSchroe}) normalized to be $1$ at some point $q_1-\varepsilon$ near $q_1$. For $E$ near $0$, the geometry Stokes curves will vary, but the Stokes zones $R$ and $R''$, figure \ref{RDRWp18}, will remain well-defined.

Let the resurgent symbol $c(E,\zeta)$ be one of the connection coefficient between the Stokes zones $R$ and $R''$ for different values of $E$ with $|E|$ small enough, given in the basis $\psi_-(q,E)$, $\psi_+(q,E)$.

 Although $c(E,\zeta)$ (or, by abuse of notation $c(E,h)$) is discontinuous with respect to $E$, the corresponding resurgent function defined for $E>0$ as $C(E,h)={\cal L}[{\bf s}_{\arg h +}c(E,\zeta)]$  and analytically continued for other values of $E$, is a holomorphic function of $E$ (for $E\in U\subset \C$), i.e. has a representative (as a sectorial germ $\mod {\cal E}^{-\infty}$) that is holomorphic with respect to $E$. This follows by the same argument as given in ~\cite[p.60, section 4.1]{DP99}. This will allow to calculate $c(E,\zeta)$ for $E>0$, pass to $C(E,h)$, analytically continue  the result to the case $E=E_rh$ and pass again to a resurgent symbol $c(E_rh,\zeta)$. 

We will obtain a formula for the hyperasymptotic expansion of $C(E,h)$ for $E$ a positive real number, but this hyperasymptotic expansion will look singular for $E\to 0$, and so there is no hope of naively subsituting $hE_r$ for $E$ in that expression. When $E\to 0$, the behavior of the microfunction decomposition of $C(E,h)$ resembles the behavior of a WKB asymptotic wave function near a turning point. Namely,  $C(E,h)$ is a resurgent function of $h$ for any $E$, but when $E\in \C\backslash\{0\}$ and orbits around the origin, the decomposition of $C(E,h)$ into microfunction undergoes Stokes phenomena, as we shall see shortly. When one analyzes the behavior of a WKB wave function near a simple turning point, one compares it to the Airy function; we will see that $C(E,h)$ for $E\to 0$ should be compared with an expression involving Gamma function. 

{\bf Notation}. Following ~\cite{DP99} and others, denote $a^{\rho}=e^{2\pi i s_{\rho}}$ the monodromy of a formal WKB solution along a contour (or even any path) $\rho$ on the Riemann surface of the classical momentum.

\subsection{Connection paths for nonsingular $E$} \label{ConnPathNonsingE}

Let us assume $0<E<<1$ and study the ``deformed" equation (\ref{DeformedSchroe}),
its Stokes pattern and its connection problem.

\subsubsection{Case when $f$ has a local minimum}

\begin{figure}
\includegraphics{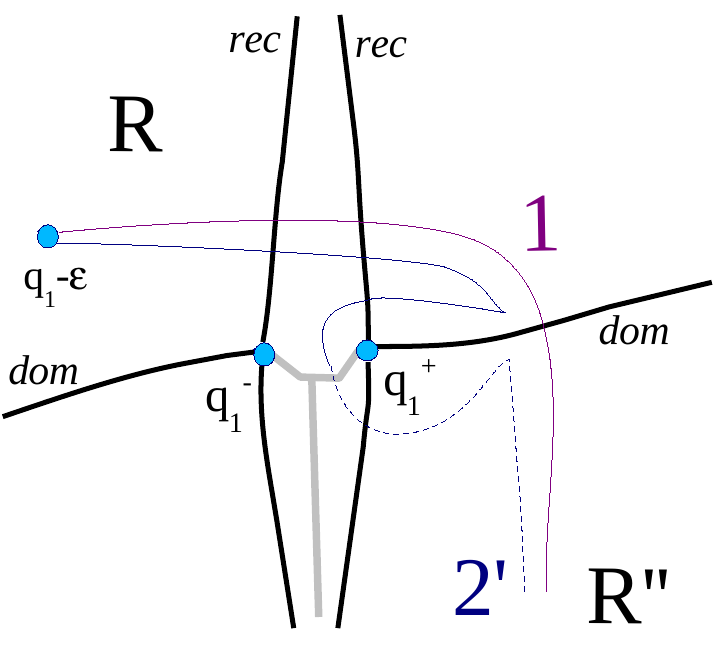} \includegraphics{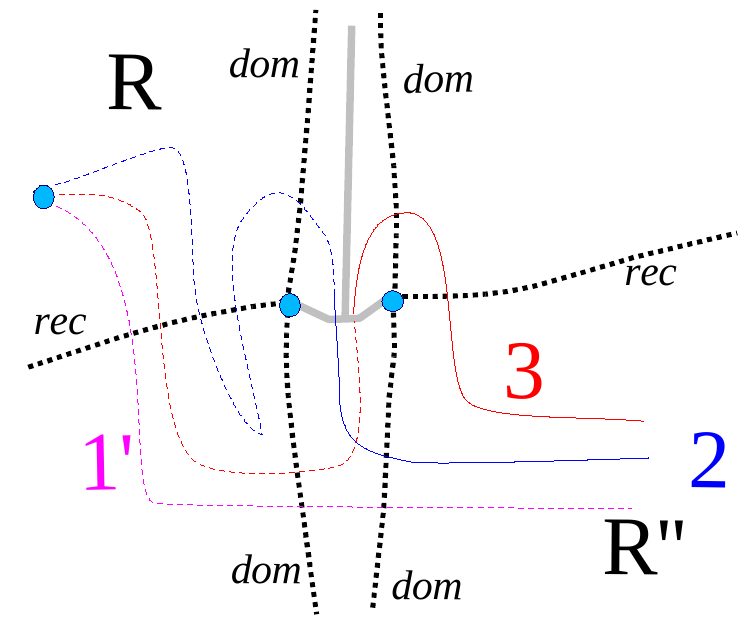} \caption{The Stokes pattern for (\ref{DeformedSchroe}) near $q_1$ and the {\it ``connection paths''} on the first sheet (left) and on the second sheet (right).} \label{Paper2p10}
\end{figure}

Consider the complex plane of $q$ near the point $q_1$ where $f(q)$ has a local minimum. The equation (\ref{DeformedSchroe}) for $0<E<<1$ will have two turning points $q_1^-$ and $q_1^+$; the Stokes curves and the Stokes zones $R$, $R''$ are shown on fig.\ref{Paper2p10}.

Let us consider two formal solutions $\psi_+(q,h)$ and $\psi_-(q,h)$ of (\ref{DeformedSchroe}) corresponding to the first and to the second sheets of the Riemann surface of the classical momentum and normalized in such a way that $\psi_+(q_1-\varepsilon)=\psi_-(q_1-\varepsilon)=1$. In order to make them univalued functions of $q$, introduce additional ``vertical'' cuts on both sheets as shown on fig.\ref{Paper2p10}.

For $E\to 0$ the formal solution $\psi_+$, resp. $\psi_-$, will behave like $e^{\frac{f(q)-f(q_1-\varepsilon)}{h}}$, resp. $e^{-\frac{f(q)-f(q_1-\varepsilon)}{h}}$, and so $\psi_+$ and $\psi_-$ will be dominant (exponentially growing) or recessive (exponentially decreasing)  in the direction away from the turning point 
along the Stokes curves marked by {\it dom} or {\it rec} on fig.\ref{Paper2p10}.


Following ~\cite{DDP97} and using  Theorem \ref{ThmVoros}, an actual solution representable by a formal WKB solution $\psi_+$ in the Stokes zone $R$ is representable in the zone $R''$ as the sum of analytic continuations of that formal solution along the paths $1$ and $2'$, fig.\ref{Paper2p10}, left, and the path $2'$ is homotopic to the path $1'$ of analytic continuation of the formal solution $\psi_-$ from $R$ to $R''$ concatinated with $\sigma_1$.
  

Similarly, an actual solution representable by the formal solution $\psi_-$ in $R$ is representable by the sum of analytic continuations of $\psi_-$ along the paths $1'$, $2$, $3$, fig.\ref{Paper2p10}, right, and the paths $2$ and $3$ are homotopic to the path $1$ of analytic continuation of the formal solution $\psi_+$ from $R$ to $R''$ concatinated with the path reverse to $\sigma_1$ and with the path $\sigma''_1$, respectively.

Paths $1$, $1'$, $2$, $2'$, $3$ are called connection paths in the literature on resurgent analysis. Paths $\gamma_j$, $\sigma_j$, etc. have been defined on fig.\ref{MonodromyExp}, \ref{nuthesisp8}, and \ref{Paper2p12}.


We see that 
$$C^{(q_1)} \  = \  \left( \begin{array}{cc} c_{++} & c_{-+} \\ c_{+-} & c_{--} \end{array} \right) \ 
= \ \left( \begin{array}{cc} 1 & (a^{\sigma_1})^{-1} + a^{\sigma''_1} \\ a^{\sigma_1}  & 1 \end{array} \right) \ = \ $$ 
$$ \ = \ \left( \begin{array}{cc} 1 & (a^{\sigma_1})^{-1}\left( 1 +e^{\delta'_1} \right)  \\ a^{\sigma_1}  & 1 \end{array} \right) \ = \  \left( \begin{array}{cc} 1 & (a^{\sigma_1})^{-1}\left( 1 - a^{\gamma_1} \right)  \\ a^{\sigma_1}  & 1 \end{array} \right), $$
where we have used (\ref{SigmaBis}).


The monodromies $a^{\delta_1}, a^{\delta'_1}$ are analytic in $E$ when $E\to 0$, the limiting behavior of $s_{\sigma_1}$ will be studied in more details the next subsection.

\subsubsection{Case when $f$ has a local maximum.} Around $q_2$ which splits for $E>0$ into two simple turning points $q_2^{\pm}$, the horizontal Stokes curves on the first sheet will be recessive.

\begin{figure}
\includegraphics{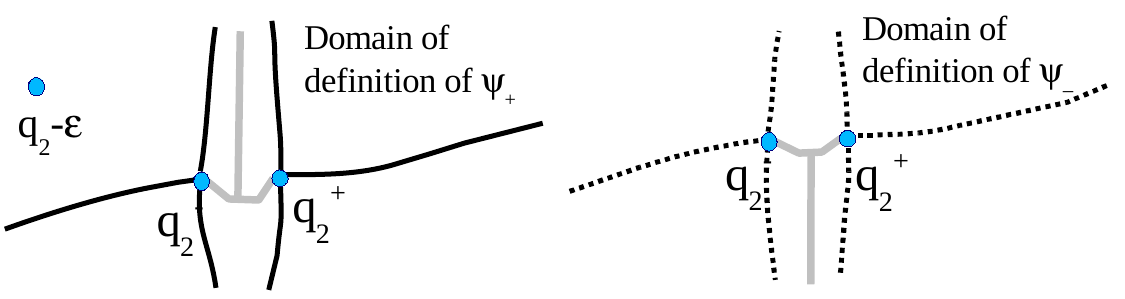} \caption{Domain of definition of $\psi_+$ (left) and $\psi_-$ (right) near $q_2$} \label{Paper2p13}
\end{figure}

Let us consider two formal solutions $\psi_+(q,h)$ and $\psi_-(q,h)$ of (\ref{DeformedSchroe}) corresponding to the first and to the second sheets of the Riemann surface of the classical momentum and normalized in such a way that $\psi_+(q_2-\varepsilon)=\psi_-(q_2-\varepsilon)=1$. In order to make them univalued functions of $q$, introduce additional ``vertical'' cuts on both sheets as shown on fig.\ref{Paper2p13}.

In all of the considerations of the $q_1$ case reverse the roles of the upper and the lower sheets and obtain:
$$C^{(q_2)} \ = \ \left( \begin{array}{cc} 1 &  a^{\sigma'_2} \\ (a^{\sigma'_2})^{-1} + a^{{\bar\sigma}_2}  & 1 \end{array} \right) \ = \ 
\left( \begin{array}{cc} 1 &  a^{\sigma'_2} \\ (a^{\sigma'_2})^{-1}(1 + a^{\delta_2} ) & 1 \end{array} \right) $$
$$ \ = \ \left( \begin{array}{cc} 1 &  a^{\sigma'_2} \\ (a^{\sigma'_2})^{-1}(1 - (a^{\gamma_2})^{-1} ) & 1 \end{array} \right). $$



\subsection{Asymptotic representation of $C$} \label{AsymptRepresOfC}

Throughout this subsection $E$ is a number $0<|E|<<1$ such that all turning points are simple. Let $q_1^{-}(E)$ and $q_1^{+}(E)$ be simple turning points near $q_1$. 

Earlier we have have associated to a connection coefficient  $c(E,h):=c_{+-}(E,h)=\exp (2\pi i s_{\sigma_1})$ (which is a resurgent symbol) a resurgent function $C(E,h)$ that is representable by $c(E,h)$ for $E>0$ and has (as a class $\mod {\cal E}^{-\infty}$) a representative holomorphic with respect to $E$ in a full complex neighborhood of the origin.



{\bf Definition.} (cf. ~\cite[def.0.6.3]{DP99}) A parameter-dependent resurgent function $\varphi(u,h)$ is said to {\bf depend regularly} on $u$ near $u_0$ its major has the origin as its only singularity in a small disc $|\xi|<\varepsilon$, independent of $u$ near $u_0$. 

Let, further, $\Phi(u,\xi)$ be a major of a parameter dependent resurgent function $\varphi(u,h)$. As $u\to u_0$,  some singularities on the Riemann surface of $\Phi(u,\xi)$, say, $\xi=s_1(u)$ and $\xi=s_2(u)$ may collide, $s_1(u_0)=s_2(u_0)$, in which case they are called {\bf confluent singularities} for $u\to u_0$, as opposed to other singularities which are {\bf isolated} for $u\to u_0$.

A parameter-dependent microfunction centered at $\omega(u)$ is called a {\bf local resurgence constant}, if it has a representative ${\pmb \varphi}(u,\xi)$ without singularities on the first sheet confluent with $\omega(u)$ for $u\to u_0$.  Equivalently, we require that $\dot \Delta_\omega {\pmb \varphi} =0$ for every $\omega$ in some neighborhood of $u_0$ independent of $u$.

These concepts apply to our situation with $u$ equal to $E$ and $u_0=0$.

\begin{Thm} (Compare ~\cite[Th.4.1.1, p.61]{DP99}.) We have 
\begin{equation} C(E,h) = e^{\frac{2[f(q_1)-f(q_1-\varepsilon)]}{h}} \frac{\sqrt{2\pi}h^{-s_{\delta_1}}}{\Gamma\left(s_{\delta_1}+\frac{1}{2}\right)} C^{red}(E,h), \label{EqTh441} \end{equation}
where $s_{\delta_1}=s_{\delta_1}(E,h)$ is the monodromy exponent along $\delta_1$, whereas $C^{red}(E,h)$  is an invertible holomorphic function of $E$, resurgent with respect to $h$ and  corresponding to an elementary simple resurgent symbol, depending regularly on $E$ in a neighborhood of $0$. \end{Thm} 


{\bf Remark.} One can see from the integral representations given in  ~\cite{Bo94} that the function $\frac{h^{-s_{\delta_1}}}{\Gamma(s_{\delta_1}+\frac{1}{2})}$ is resurgent with respect to $h$ and has a representative (as a class modulo ${\cal E}^{-\infty}$) that is holomorphic with respect to $E$. However, the function $h^{-s_{\delta_1}}$ and $\Gamma(s_{\delta_1}+\frac{1}{2})$ are not resurgent by themselves because they do not satisfy the growth conditions for $h\to 0+$.

A more precise calculation of  $C^{red}$ is carried out below by using the exact matching method.

\textsc{Proof of the theorem.} 



In the subsection \ref{Alienq1} we will give a proof (compare ~\cite[p.58]{DP99}) 
%
%
that  $c$ satisfies the following local resurgence equation (see section \ref{HomDePassage} for notation) 
\footnote{
This equation is a local resurgence equation in the following sense: the statement about the alien derivative should hold for $|E|$ small enough (how small, depends on $n$), and the equality is true only modulo microfunctions whose supports are separated from zero for $E\to 0$.
}
\begin{equation} \dot\Delta_{-in\omega_{\gamma_1}} c = \frac{(-1)^{n-1}}{n} a^{-n\delta_1}c \ \ \ \ (n\in\Z\backslash\{ 0\}) \label{DeltaNOmega} \end{equation}
and its minor has no other confluent singularities than the integral multiples of $\omega_{\gamma_1}=\int_{\gamma_1} i\sqrt{E-[f'(q)]^2} dq$. 

To write a general solution of that resurgence equation, recall that ~\cite[Pro I.4.3]{CNP} states (up to  rescaling of $x=1/h$) that if for a microfunction $a$
$$ {\dot\Delta}_{2\pi in} a = \frac{(-1)^{n-1}}{n} e^{-2\pi i n x}a \ \ \ \ \ \text{for} \ n\in \Z^*=\Z\backslash\{0\},$$
$$ \Delta_{\omega} a = 0 \ \ \ \ \  \  \text{for} \ \omega\not\in  2\pi i\Z^* , $$
then 
\begin{equation} ({\bf s} a)(x) = (x/e)^{x}\sqrt{2\pi}/\Gamma(x+1/2) ({\bf s} b),  \label{eq3} \end{equation}
where ${\dot\Delta}_\omega b=0$ for all $\omega\in \C$ close enough to the center of the microfunction $b$. 
Note also ~\cite{Bo94} for an excellent exposition of resurgence properties of the gamma function.

Take $s_{\delta_1}$ as a new resurgent variable instead of $x=1/h$; then the resurgence equation found earlier becomes
$$ {\dot \Delta}^{(s_{\delta_1})}_{2\pi i n}c = \frac{(-1)^{n-1}}{n} e^{-2\pi i n s_{\delta_1}} c, $$
hence, by \eqref{eq3},
$$ ({\bf s}c)(E,h) = (s_{\delta_1} e^{-s_{\delta_1}}) \frac{\sqrt{2\pi}}{\Gamma(s_{\delta_1}+\frac{1}{2})} {\bf s}(c'(E,h)), $$
for a local resurgence constant $c'$. 

Since $(hs_{\delta_1})^{-s_{\delta_1}}$ is a local resurgence constant  as well, we can absorb it in $C'$ and obtain a new local resurgent consant $C^{red}$ whose resurgent symbol is $c^{red}(E,h)$: \begin{equation} C(E,h) \ = \  \frac{\sqrt{2\pi} h^{-s_{\delta_1}}}{\Gamma(s_{\delta_1}+\frac{1}{2})} 
{\bf s}(c^{red}(E,h))  \label{eq11} \end{equation}
The facts that $C^{red}$ is holomorphic in $E$ and invertible and that the symbol $c^{red}$ is elementary and simple for all $0\ne E\in \C$ can be shown as in ~\cite{DP99}.
$\Box$

The contributions to the connection coefficients $C(E,h)$ due to the turning points far away are of the exponential order negative the length of some double Stokes curves for appropriate $\arg h$. This is what some authors probably mean when they say that different turning points are ``decoupled''. \\



It is clear that analogous statements will hold for the formal monodromy along $\sigma'_2$.



\subsection{Calculation of $C^{\rm red}$ by the exact matching method.} 

In this section we will take the formula \eqref{EqTh441} for the connection coefficient and pass to the limit when $E$ stops being a positive real number and becomes $hE_r$. We will use the fact that $hE_r$ can be substituted into the fraction  $\frac{h^{-s_{\delta_1}}}{\Gamma(s_{\delta_1}+\frac{1}{2})}$ easily. Thus, the real task will be to calculate $c^{red}(E,h)$ for $E$ replaced by $hE_r$.

We are studying the connection problem across the double turning point $q_k$, $k=1,2,...,2n$, for the differential equation 
\begin{equation} [-h^2 \partial_q^2 + V(q)+hV_1(q)]\psi=hE_r\psi, \label{Unshifted} \end{equation}
where in our case $V(q)=[f'(q)]^2$ and $V_1(q)=-f''(q)$.


 The exact matching method given in ~\cite[\S 5.1.1, p.74]{DP99} for the Schr\"odinger equation without the $hV_1(q)$ term and the argument justifying it applies equally well to our case. It would be perhaps desirable to give a more detailed treatment of this method and its proof, as we plan to do elsewhere.

For odd $j$, we will denote by $c'_j$  the limit of the monodromy 
$ e^{2\pi i s_{\sigma_j}}$ of formal solutions of \eqref{DeformedSchroe} when a positive real number $E$ is replaced by $hE_r$, and for even $j$ let  $c_j$ denote the corresponding limit of $e^{2\pi i s_{\sigma_j}}$.

To simplify notation, we are going to treat two representative double turning points $q_1$ and $q_2$, where $f(q)$ has a local minimum and a local maximum, respectively; the analogous formulas will hold for other $q_j$ as well.

\subsubsection{Exact matching method around $q_1$.} \label{ExactMatchingq1}

Instead of calculating the monodromies of formal solutions of \eqref{Unshifted}, we will calculate them for the equation 
\begin{equation} [-h^2 \partial_q^2 + V(q)+h(V_1(q)-V_1(q_1))] \psi = h(E_r-V_1(q_1))\psi  = hE'_r\psi \label{ShiftedEq} \end{equation}
and then substitute $E'_r=E_r-V_1(q)$ in the answer.  Denote by ${\tilde c}^{red}$, ${\tilde s}_{\delta_1}$ etc. the objects defined from the equation \eqref{ShiftedEq} similarly to $c^{red}$, $s_{\delta_1}$ etc. for \eqref{Unshifted}.

We will calculate the asymptotic expansion in $h$ representing ${\tilde c}^{red}(E,h)$ for $0<E<<1$ from the formula
$$ {\bf s}({\tilde c}^{red}(E,h)) = \exp (2\pi i {\tilde s}_{\delta_1} )  \left[ \frac{\sqrt{2\pi} h^{-{\tilde s}_{\delta_1}}}{\Gamma({\tilde s}_{\delta_1}+\frac{1}{2})} \right]^{-1}, $$
and then substitute $hE'_r$ for $E$ into this power series. 

Easy algebra shows that 
$$ \exp (2\pi i {\tilde s}_{\sigma_1} )  \ = \ -i\exp\left( \frac{i}{h}\Delta S \ + \ \Delta R \ + \ O(h) \right), $$
where 
$$\Delta S (E,q_0) \ = \ \int_{\sigma_1} \sqrt{E-(f')^2} dq', \ \ \ \Delta R = \int_{\sigma_1} \frac{f''-f''(q_1)}{2i\sqrt{E-(f')^2}}  dq$$
with determinations of the square root given as on figure \ref{nuthesisp5}.

The Stirling formula (applied for $|h|\to 0$ with $E$ fixed) together with the property ${\tilde s}_{\delta_1}+\frac{1}{2}= -{\tilde s}_{\gamma_1}$ and the definition $\omega_{\gamma_1}=\int_{\gamma_1} \sqrt{E-[f'(q)]^2} dq$ yield
$$\frac{\sqrt{2\pi} h^{-{\tilde s}_{\delta_1}}}{\Gamma({\tilde s}_{\delta_1}+\frac{1}{2})} \ = \ \exp \left( \frac{1}{h} (-\frac{\omega_{\gamma_1}}{2\pi} + \frac{\omega_{\gamma_1}}{2\pi } \ln [-\frac{\omega_{\gamma_1}}{2\pi }] ) \right) (1+O(h)). $$
Therefore 
$$ {\tilde c}^{red}(E,h) = -i \exp\left\{ \frac{1}{h}\left[i\Delta S + \frac{\omega_{\gamma_1}}{2\pi} - \frac{\omega_{\gamma_1}}{2\pi } \ln [-\frac{\omega_{\gamma_1}}{2\pi }\right] + \Delta R \right\} (1+O(h)). $$

An elementary but lengthy calculation shows that $i\Delta S + \frac{\omega_{\gamma_1}}{2\pi} - \frac{\omega_{\gamma_1}}{2\pi } \ln [-\frac{\omega_{\gamma_1}}{2\pi }]$ is analytic with respect to $E$ for $E$ near zero, and 
$$ i \Delta S + \frac{\omega_{\gamma_1}}{2\pi} - \frac{\omega_{\gamma_1}}{2\pi } \ln [-\frac{\omega_{\gamma_1}}{2\pi }] \ = \ \ \ \ \ $$
$$ \ \ \ = \ 2[f(q_1)-f(q_1-\varepsilon)] - \int_{\sigma_1} \frac{Edq}{2i\sqrt{E-(f')^2}} - \frac{\omega_{\gamma_1}}{2\pi } \ln [-\frac{\omega_{\gamma_1}}{2\pi } ]+ O(E^2) .$$

Using results of section \ref{MonodrExp}, we have 
  $$ \int_{\sigma_1} \frac{dq}{2i\sqrt{E-(f')^2}} \ = \  \frac{1}{f''(q_1)} \int_{\sigma_1} \frac{f''(q) dq}{2i\sqrt{E-(f')^2}} \ - \ \frac{1}{f''(q_1)} \int_{\sigma_1} \frac{f''(q)-f''(q_1)}{2i\sqrt{E-(f')^2}}dq \ = \ $$
$$ \ = \ -\frac{1}{f''(q_1)} \Ln [ -\frac{2f'(q_1-\varepsilon)}{\sqrt{E}}(1+O(E)) ] \ - \  \frac{1}{f''(q_1)} \Delta R, $$ 
and also 
$$  -E \int_{\sigma_1} \frac{dq}{2i\sqrt{E-(f')^2}} \ - \  \frac{ \omega_{\gamma_1}}{2\pi } \ln [-\frac{\omega_\gamma(E)}{2\pi}] \ = \ $$
\small
$$ \ = \  E \left\{ \frac{1}{f''(q_1)} \Ln [-\frac{2f'(q_1-\varepsilon)}{\sqrt{E}} ] \ + \ \frac{1}{2f''(q_1)}  \ln \left( \frac{E}{2f''(q_1)}\right)  \right\} + o(E) \ = \ $$ \normalsize
$$ \ = \ \frac{E}{f''(q_1)} \Ln [ -\frac{2f'(q_1-\varepsilon)}{ \sqrt{2f''(q_1)}}] \ + \ o(E). $$

Now we can substitute $E=hE'_r$ in the above expression and calculate  the connection coefficient for the equation \eqref{ShiftedEq} (up to a factor $(1+O(h)+O(E))$) 
$$ \left. e^{2\pi i {\tilde s}_{\delta_1}} \right|_{E=hE'_r}  \ \approx \ -i e^{\frac{2[f(q_1)-f(q_1-\varepsilon)]}{h}} \frac{\sqrt{2\pi}h^{-\frac{E'_r}{2f''(q_1)}}}{\Gamma\left(\frac{E'_r}{2f''(q_1)}+\frac{1}{2}\right)} \times \ \ \ \ \ $$
$$ \ \ \ \ \times 
\exp \left[ \frac{E'_r}{f''(q_1)}\Ln [ -\frac{2f'(q_1-\varepsilon)}{ \sqrt{2f''(q_1)}}] \ + \ \left( \frac{E'_r}{f''(q_1)}-1\right)\Delta R \right], $$ 
where it is important that $\Delta R$ as a function of $E$ is bounded for $E\to 0$.
Inserting  $E'_r=E_r+f''(q_1)$,   obtain:
$$ \left. e^{2\pi i {s}_{\delta_1}} \right|_{E=hE_r} \ \approx \ -i \frac{\sqrt{2\pi}h^{-\frac{E_r+f''(q_1)}{2f''(q_1)}}}{\Gamma\left(\frac{E_r+f''(q_1)}{2f''(q_1)}+\frac{1}{2}\right)}  \exp \left[ \frac{E_r+f''(q_1)}{f''(q_1)}\Ln [ -\frac{2f'(q_1-\varepsilon)}{ \sqrt{2f''(q_1)}}] \right] \times $$ $$ \times \exp \left[ \left( \frac{E_r}{f''(q_1)}\right) \Delta R  \ + \ \frac{2[f(q_1)-f(q_1-\varepsilon)]}{h} \right]. $$
We will need this result for $E_r$ not a complex number, but a resurgent function of negative exponential type, in which case we can simplify: 
$$ c'_1 |_{E_r=O(h)} \ = \ \left. e^{2\pi i {s}_{\delta_1}} \right|_{E=hE_r=O(h^2)}  \approx \ 
\frac{2i\sqrt{\pi}f'(q_1-\varepsilon)}{\sqrt{hf''(q_1)}}  e^{\frac{2[f(q_1)-\Phi(q_1-\varepsilon)]}{h}}, $$
where $\approx$ means that an equality holds up to a factor $1+O(h)$.


\subsubsection{Exact matching method around $q_2$.} \label{ExactMatchingq2}  Following the same line of thought for the situation around the turning point $q_2$, we conclude
$$ c_2 \ \approx 
  - E_r \frac{ i \sqrt{\pi h} }{2\sqrt{|f''(q_2)|}f'(q_2-\varepsilon)}  e^{-\frac{2[f(q_2)-f(q_2-\varepsilon)]}{h}}, $$
where $\approx$ stand for an equality up to a factor $(1+O(h))$.


\subsection{Alien differential equation for the connection coefficient.} \label{Alienq1}

We will derive here an alien differential equation of a connection coefficient $c=e^{2\pi i s_{\sigma_1}}$, where $q_1\in \R$ is a local minimum of the function $f$. Up to small details, a similar argument can be repeated for a local maximum of $f$ as well.

The main theoretical ingredient is the following:

\begin{Thm} (\cite{DDP93}) Let $\gamma$ be a path on the Riemann surface of the momentum (closed or not). Let $a^{\gamma}$  the monodromy of the formal solution along $\gamma$.  Fix a resummation direction, or $\arg h$, to be $\alpha$. Then:\\
1) If $\gamma$ intersects no Stokes curves in direction $\alpha$, then $\Delta_\alpha  a^{\gamma} = 0$; \\
2) same if $\gamma$ intersects a simple Stokes curve; \\
3) if $\gamma$ intersects a double Stokes curve in the direction $\alpha$ with a period cycle $\rho$, then $\bigsigma_\alpha a^{\gamma} = (1\pm a^{\rho})a^{\gamma}$. \end{Thm}

\begin{figure} \includegraphics{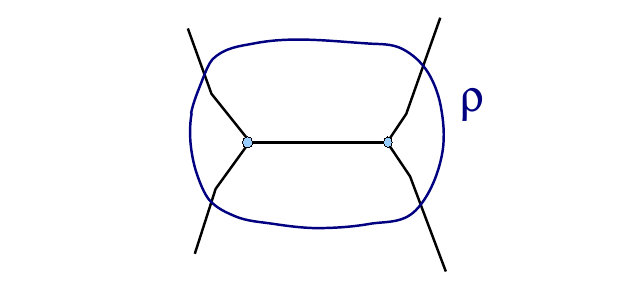} \caption{A period cycle of a double Stokes curve} \label{dSc}
\end{figure}

The period cycle of a double Stokes curve is a closed path $\rho$ encircling both ends of the double Stokes curve and oriented in such a way that the exponential type of $a^{\rho}$ is negative, figure \ref{dSc}.

\begin{figure}[h]
\includegraphics{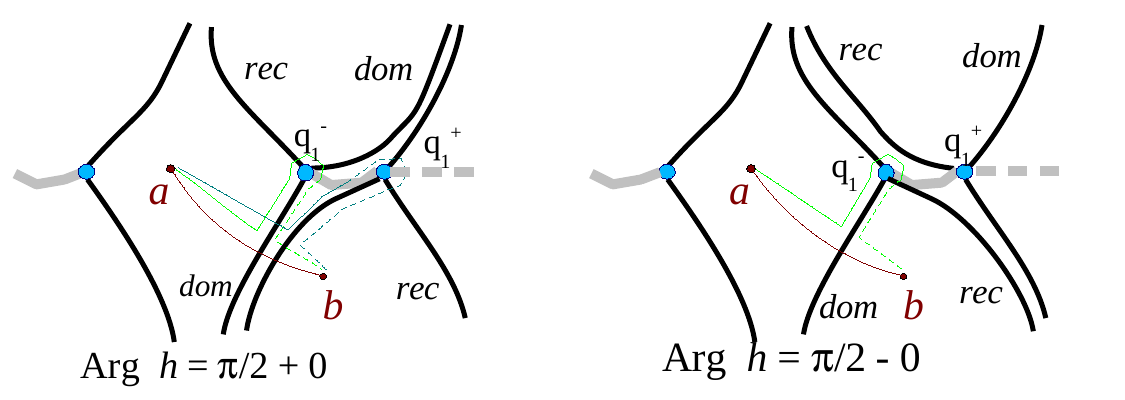} \caption{Bifurcation of the Stokes pattern around $q_1$ for $\Arg h = \frac{\pi}{2} \pm 0$.} \label{nuthesisp1}
\end{figure}

\begin{figure}[h]\includegraphics{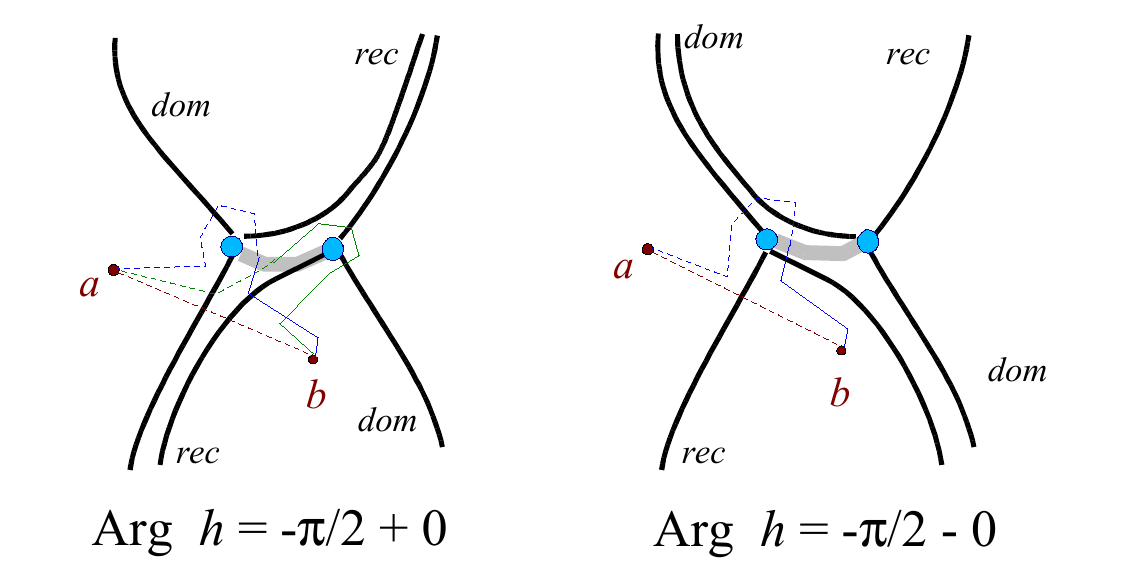} \caption{Bifurcation of the Stokes pattern around $q_1$ for $\Arg h = -\frac{\pi}{2}\pm 0$} \label{nuthesisp17}
\end{figure}

We will consider the equation \eqref{DeformedSchroe} for $0<E<<1$, but for all possible values of $\arg h$. As we know, the Stokes curves depend on the resummation direction $\arg h$. In particular, for $\arg h=\pm\pi/2$ there will be a short double Stokes curve connecting $q_1^-$ and $q_1^+$, and for $\arg h$ close to these two values the Stokes pattern will ``bifurcate" as shown on figures \ref{nuthesisp1} and \ref{nuthesisp17}.

Consider the connection problem for resurgent solutions of \eqref{DeformedSchroe} from a neighborhood of the point $a$ to the neighborhood of $b$ shown on fig. \ref{nuthesisp1}. Let  $\psi_+$ and $\psi_-$ be, as usual, formal WKB solutions such that $\psi_+(a)=\psi_-(a)=1$. They will be univalued functions of $q$ once we make a cut between $q_1^-$ and $q_1^+$ and a horizontal cut (dashed line) to the right of $q_1^+$. 

\begin{figure}[h]\includegraphics{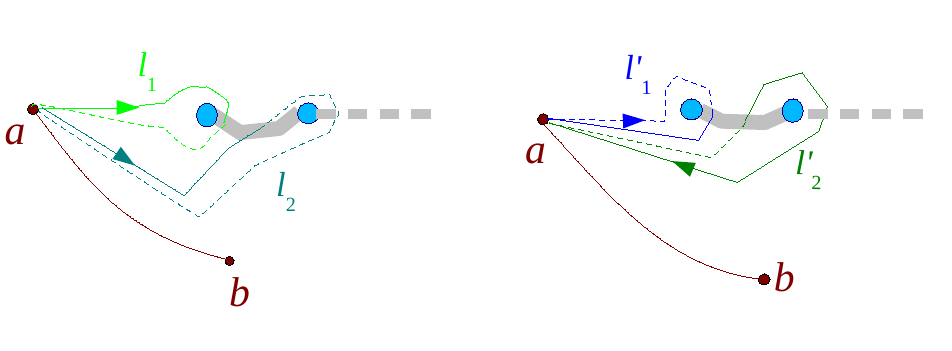}  \caption{Connection paths $\ell_1$, $\ell_2$, $\ell'_1$, $\ell'_2$.} \label{nuthesisp16}
\end{figure}

For $\Arg h =  \pi/2 + 0$ the actual solution of \eqref{DeformedSchroe} represented by $\psi_+$ at $a$ is represented by $\psi_+ + (a^{\ell_1} + a^{\ell_2})\psi_-$ at b; for $\Arg h = \pi/2 - 0$ the representation at $b$ is $\psi_+ + (a^{\ell_1})\psi_-$. Since $\psi_-(b)$ is a resurgence constant, we obtain
$$ {\bf s}_{\pi/2+0} ((a^{\ell_1} + a^{\ell_2})) = {\bf s}_{\pi/2-0} a^{\ell_1},  $$
$$ {\bigsigma}_{\pi/2} (a^{\ell_1} + a^{\ell_2}) = a^{\ell_1}, $$
and since  $a^{\ell_2} (a^{\ell_1})^{-1} = a^{\delta_1}$ by deformation of the integration contour, we obtain:
$$ {\bigsigma}_{\pi/2} a^{\ell_1} \ = \ a^{\ell_1} (1+{\bigsigma}_{\pi/2} a^{\delta_1})^{-1} $$
Using the definition of the alien derivative as $\Ln(\bigsigma_\alpha)$, obtain \eqref{DeltaNOmega} for positive $n$. 

Now let us study a similar bifurcation of the Stokes pattern for $\Arg h = -\frac{\pi}{2}\pm 0$.

For $\Arg h =  -\pi/2 + 0$ the solution represented by $\psi_+$ at $a$ is represented by $\psi_+ + (a^{\ell'_1} + a^{\ell'_2})\psi_-$ at $b$; for $\Arg h = -\pi/2 - 0$ the representation at $b$ is $\psi_+ + (a^{\ell'_1})\psi_-$.  Since $\psi_+(b)$ is a resurgence constant, therefore 
$$ {\bf s}_{-\pi/2+0} (a^{\ell'_1} + a^{\ell'_2}) = {\bf s}_{-\pi/2-0} a^{\ell'_1} \ ;  $$
$$ {\bigsigma}_{-\pi/2} (a^{\ell'_1} + a^{\ell'_2}) =  a^{\ell'_1} .  $$
Using the fact that ${\bigsigma}$ is a multiplicative homomorphism 
get:
$$ {\bigsigma}_{-\pi/2} (a^{\ell_1}(a^{\ell'_1} + a^{\ell'_2})) =  a^{\ell'_1} \bigsigma_{-\pi/2} a^{\ell_1} \ ;  $$
$$ (a^{\ell'_1})^{-1}{\bigsigma}_{-\pi/2} (-1  + a^{\ell_1}a^{\ell'_2})) =   \bigsigma_{-\pi/2} a^{\ell_1} \ ;  $$
$$ a^{\ell_1}  {\bigsigma}_{-\pi/2} (1  -  (a^{\gamma_1})^{-1}) =   \bigsigma_{-\pi/2} a^{\ell_1} .  $$
We will now use the equality $\bigsigma_{-\pi/2} (1-(a^{\gamma_1})^{-1})= 1- (a^{\gamma_1})^{-1} + r$, where the exponential type of $r$ is estimated by the canonical length of the Stokes curves for $\arg h=\pi/2$ starting at $q_1^{\pm}$. Modulo terms of that exponential type, obtain:  
$$ \bigsigma_{-\pi/2} a^{\ell_1} \approx a^{\ell_1} (1+a^{\delta_1}), $$
and taking the logarithm of $\bigsigma_\alpha$, obtain the part of $\eqref{DeltaNOmega}$ for negative $n$.

\section{Transfer matrix and quantization condition.} \label{TMQC}

In this section we are studying the quantization condition for the Witten Laplacian with the superpotential $f$ having $n$ local minima and $n$ local maxima. The eigenvalue of the Witten Laplacian in this section will be written as $hE_r$, $E_r\in \C$.

\begin{figure}[h]\includegraphics{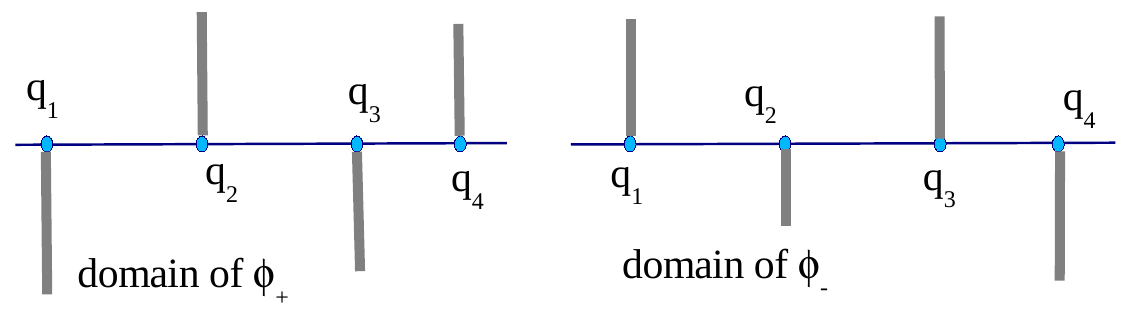} \caption{Domains of $\phi_+$ and $\phi_-$.} \label{nuthesisp9}
\end{figure}

Let $\phi_+,\phi_-$ be the formal resurgent solutions of 
\begin{equation} (-h^2 \partial^2_x + [f']^2 - hf'' )\phi = h E_r \phi \label{ReducedSchroe} \end{equation}
corresponding to the first and second sheet of the Riemann surface, normalized in such a way that $\psi_{+}(q_0)=\psi_{-}(q_0)=1$ and defined on the domains (complex plane with vertical cuts starting at $q_j$ ) shown on fig.\ref{nuthesisp9}. 

\subsection{The transfer matrix.} \label{TransMtx} 

We would like to write down a condition that for a given $E_r$ our equation \eqref{ReducedSchroe} 
has a periodic solution, $\varphi(q,h)=\varphi(q+1,h)$. After that, the eigenvalue problem for (\ref{ReducedSchroe}) will reduce to solving that condition with respect to $E_r$.

The following definition of the {\bf transfer matrix} $F$ will in spirit resemble the definition of the connection matrices between two Stokes zones. Suppose an actual solution $\varphi(q,h)$
is representable as $A_- \psi_-+A_+\psi_+$ around $q_0$, where $A_-$ and $A_+$ are some constant resurgent symbols and $\psi_-$ and $\psi_+$ the elementary formal solutions of our differential equation with, to fix ideas,  $\psi_-(q_0)=\psi_+(q_0)=1$, and, as we assume the statements of section \ref{ExistProb}, $q_0\in \C$ can be chosen arbitrarily so that $f'(q_0)\ne 0$. Since the coefficients of \eqref{ReducedSchroe} are periodic, the function $\varphi(q+1)$ will also be its solution, and therefore representable as $B_{-}\psi_{-}+B_{+}\psi_{+}$. We will define the transfer matrix $F$ by the relation
$$ \left(\begin{array}{cc} B_- \\ B_+ \end{array}\right) \ = \ F  \left(\begin{array}{cc} B_- \\ B_+ \end{array}\right).$$ 

This matrix $F$ is obtained as a composition of analytic continuations of formal solutions inside the Stokes regions and connection matrices between different Stokes regions, as explained below.

The entries of the matrix $F$ are formal resurgent symbols dependent on $E_r$. It is clear that $E_r$ is an eigenvalue of the Witten Laplacian if and only if it is a solution of the following {\bf quantization condition}:
\begin{equation} \det ( F(E_r) - Id ) \ = \ 0. \label{qCond81} \end{equation}
In fact, we will presume that $E_r$ is a number to set up this equation, then solve it and find its resurgent solutions, then by ~\cite{G1} one can substitute $E_r$ into the original equation and obtain its resurgent solutions satisfying the periodic boundary conditions.


\subsection{From connection matrices to quantization condition} \label{ConnMatQuantCond}

If in the basis $\psi_+,\psi_-$ of formal WKB solutions such that $\psi_+(q_1-\varepsilon)=\psi_-(q_1-\varepsilon)=1$ the connection matrix across the turning point $q_1$ equals $C$, then in the basis $\phi_+,\phi_-$ of formal WKB solutions such that $\phi_+(q_0)=\phi_-(q_0)=1$ the corresponding matrix will be written as 
$$ \left( \begin{array}{cc} \phi_+(q_1-\varepsilon) & 0 \\ 0 & \phi_-(q_1-\varepsilon) \end{array} \right)^{-1} C  
\left( \begin{array}{cc} \phi_+(q_1-\varepsilon) & 0 \\ 0 & \phi_-(q_1-\varepsilon) \end{array} \right) . $$
Composing connection matrices from the Stokes zone containing  $q_1-\varepsilon$ (cf. fig. 
\ref{nuthesisp10}) to the zone containing $q_2-\varepsilon$, from the zone containing $q_2-\varepsilon$ to the zone containing $q_3-\varepsilon$, etc, to the zone containing  $(q_1-\varepsilon)+1$, obtain:

\begin{Prop} \label{PropStructureF} In the basis $\phi_+,\phi_-$ the transfer matrix $F$ equals
$$ F \ = \ \left( \begin{array}{cc} \frac{\phi_+(1)}{\phi_+(q_{2n}-\varepsilon)} & 0 \\ 0 & \frac{\phi_{-}(1)}{\phi_{-}(q_{2n}-\varepsilon)} \end{array} \right) C^{(q_{2n},\varepsilon)} \left( \begin{array}{cc} \frac{\phi_+(q_{2n-1}-\varepsilon)}{\phi_+(q_{2n-1}-\varepsilon)} & 0 \\ 0 & \frac{\phi_{-}(q_{2n}-\varepsilon)}{\phi_{-}(q_{2n-1}-\varepsilon)} \end{array} \right) \times \dots \ \ \ \ \ \ \ $$
$$ \ \ \ \ \ \ \dots \times
\left( \begin{array}{cc} \frac{\phi_+(q_{2}-\varepsilon)}{\phi_+(q_{1}-\varepsilon)} & 0 \\ 0 & \frac{\phi_{-}(q_{2}-\varepsilon)}{\phi_{-}(q_{1}-\varepsilon)} \end{array} \right)
  C^{(q_1,\varepsilon)} \left( \begin{array}{cc} \phi_+(q_1-\varepsilon) & 0 \\ 0 & \phi_-(q_1-\varepsilon) \end{array} \right).   $$ 
Here $C^{(q_j,\varepsilon)}$ is a connection matrix across a double turning point $q_j$ in the basis of formal WKB solutions normalized to $1$ at $q_j-\varepsilon$, as in section \ref{ConnectionFlae}.
\end{Prop}

The  product of the matrices in proposition \ref{PropStructureF} has the structure
$$ F \ = \ \left( \begin{array}{cc} B_{2n} & \\ & B'_{2n} \end{array} \right)  \left( \begin{array}{cc} 1 & c_{2n} \\ c'_{2n} & 1 \end{array} \right)  \left( \begin{array}{cc} A_{2n-1} & \\ & A'_{2n-1} \end{array} \right)  \left( \begin{array}{cc} 1 & c_{2n-1} \\ c'_{2n-1} & 1 \end{array} \right) \times \dots \times $$ $$ \ \ \ \ \ \ \ \  \times \left( \begin{array}{cc} A_2 & \\ & A'_2 \end{array} \right) \left( \begin{array}{cc} 1 & c_{2} \\ c'_{2} & 1 \end{array} \right)  \left( \begin{array}{cc} A_1 & \\ & A'_1 \end{array} \right) \left( \begin{array}{cc} 1 & c_{1} \\ c'_{1} & 1 \end{array} \right)  \left( \begin{array}{cc} B_0 & \\ & B'_0 \end{array} \right). $$

Put $$F' =  \left( \begin{array}{cc} B_0 & \\ & B'_0 \end{array} \right) F  \left( \begin{array}{cc} B_0 & \\ & B'_0 \end{array} \right)^{-1}$$
and put $A_{2n} := B_{2n}B_0$, then 
$$ F' \ = \  \left( \begin{array}{cc} A_{2n} & A_{2n} c_{2n} \\ A'_{2n} c'_{2n} & A'_{2n} \end{array} \right)   \left( \begin{array}{cc} A_{2n-1} & A_{2n-1} c_{2n-1} \\ A'_{2n-1} c'_{2n-1} & A'_{2n-1} \end{array} \right)   \times \dots \times $$ $$ \ \ \ \ \ \ \ \  \times  \left( \begin{array}{cc} A_2 & A_2 c_{2} \\ A'_2 c'_{2} & A'_2 \end{array} \right)   \left( \begin{array}{cc} A_1 & A_1 c_{1} \\ A'_1 c'_{1} & A'_1 \end{array} \right),   $$
$$ F' \ = \   \left( \begin{array}{cc} A_{2n} A_{2n-1} + A_{2n} c_{2n} A'_{2n-1} c'_{2n-1} & A_{2n} A_{2n-1} c_{2n-1} + A_{2n} c_{2n} A'_{2n-1} \\ A'_{2n} c'_{2n} A_{2n-1} + A'_{2n} A'_{2n-1} c'_{2n-1} & A'_{2n} c'_{2n} A_{2n-1} c_{2n-1} + A'_{2n-1} A'_{2n} \end{array} \right)   \dots  $$ $$ \ \ \ \ \dots    \left( \begin{array}{cc} A_2 A_1 + A_2c_2 A'_1 c'_1 & A_2 A_1 c_{1} + A_2 c_2 A'_1 \\ A'_2 c'_2 A_1 + A'_2 A'_1 c'_{1} & A'_2 c'_2 A_1 c_1 + A'_1 A'_2 \end{array} \right) .  $$


Let $\mu_k = c_k c'_k$ and consider what ~\cite{DDP97} call {\it monodromies along the tunneling cycles:} 
$$ \tau_{2k-1} = c'_{2k-1} A_{2k-1}^{-1} c_{2k} A'_{2k-1} \ , \ \ \ \tau_{2k} = c_{2k} (A'_{2k})^{-1} c'_{2k+1} A_{2k}, $$
where $k=1,...,n$ and $c_{2n+1} = c_1$. 

$$ F' \ = \ \left( \begin{array}{cc} c_{2n} & \\ & 1 \end{array} \right) \times \ \ \ \ \ $$
$$ \ \times 
 \left( \begin{array}{cc} A_{2n} A_{2n-1}c_{2n}^{-1} + A_{2n}  A'_{2n-1} c'_{2n-1} & A_{2n} A_{2n-1} c_{2n}^{-1} c_{2n-1} + A_{2n}  A'_{2n-1} \\ A'_{2n}  A_{2n-1}c'_{2n} + A'_{2n} A'_{2n-1} c'_{2n-1}  & A'_{2n}  A_{2n-1} c_{2n-1} c'_{2n} + A'_{2n-1} A'_{2n}  \end{array} \right)  \times \dots  \ \ \ \ \ \ \ $$ $$ \ \ \ \ ... \times
 \left( \begin{array}{cc} c_{2} & \\ & 1 \end{array} \right) \left( \begin{array}{cc} A_2 A_1 c_2^{-1} + A_2 A'_1 c'_1 & A_2 A_1 c_{1} c_2^{-1} + A_2  A'_1 \\ A'_2  A_1 c'_2 + A'_2 A'_1 c'_{1}  & A'_2  A_1 c_1 c'_2 + A'_1 A'_2  \end{array} \right) .  $$

After some calculations, get 
\footnotesize
$$ F' \ = \ \left( \begin{array}{cc} c_{2n}A_{2n} (A'_{2n})^{-1} & \\ & 1 \end{array} \right) 
 \left( \begin{array}{cc}    \tau_{2n-1}^{-1} +   1  &    \mu_{2n-1}   \tau_{2n-1}^{-1} + 1   \\   \mu_{2n}   \tau_{2n-1}^{-1} +  1   &    \mu_{2n-1}  \mu_{2n}  \tau_{2n-1}^{-1} + 1  \end{array} \right)  \left( \begin{array}{cc} c'_{2n-1} & \\ & 1 \end{array} \right) \times \dots  \ \ \ \ \ \ \ $$ $$ \ \ \ \ \ \ \times 
 \left( \begin{array}{cc} c_{2}A_2 (A'_{2})^{-1} & \\ & 1 \end{array} \right) \left( \begin{array}{cc}     \tau_{1}^{-1} +  1  &   \mu_1   \tau_{1}^{-1} + 1 \\  \mu_2   \tau_{1}^{-1} +  1   &    \mu_1 \mu_2   \tau_{1}^{-1} + 1   \end{array} \right)  \left( \begin{array}{cc} c'_{1} & \\ & 1 \end{array} \right)  A'_1 A'_2 A'_3 ... A'_{2n-1} A'_{2n}. $$ \normalsize

Define the matrix $G$ by 
$$ \left( \begin{array}{cc} c'_{1} & \\ & 1 \end{array} \right) F' = G \left( \begin{array}{cc} c'_{1} & \\ & 1 \end{array} \right), $$
then \small
$$ G \ = \ \left( \begin{array}{cc} c'_{2n+1} c_{2n}A_{2n} (A'_{2n})^{-1} & \\ & 1 \end{array} \right) 
 \left( \begin{array}{cc}    \tau_{2n-1}^{-1} +   1  &    \mu_{2n-1}   \tau_{2n-1}^{-1} + 1   \\   \mu_{2n}   \tau_{2n-1}^{-1} +  1   &    \mu_{2n-1}  \mu_{2n}  \tau_{2n-1}^{-1} + 1  \end{array} \right)   \times \dots  \ \ \ \ \ \ \ $$ $$ \ \ \ \ \ \ \times 
 \left( \begin{array}{cc} c'_3 c_{2}A_2 (A'_{2})^{-1} & \\ & 1 \end{array} \right) \left( \begin{array}{cc}     \tau_{1}^{-1} +  1  &   \mu_1   \tau_{1}^{-1} + 1 \\  \mu_2   \tau_{1}^{-1} +  1   &    \mu_1 \mu_2   \tau_{1}^{-1} + 1   \end{array} \right)  \left( \begin{array}{cc} c'_{1} & \\ & 1 \end{array} \right)  A'_1 A'_2  ... A'_{2n}, $$ \normalsize
 $$ G \ = \ \left( \begin{array}{cc} \tau_{2n} & \\ & 1 \end{array} \right) 
 \left( \begin{array}{cc}    \tau_{2n-1}^{-1} +   1  &    \mu_{2n-1}   \tau_{2n-1}^{-1} + 1   \\   \mu_{2n}   \tau_{2n-1}^{-1} +  1   &    \mu_{2n-1}  \mu_{2n}  \tau_{2n-1}^{-1} + 1  \end{array} \right)   \times \dots  \ \ \ \ \ \ \ $$ $$ \ \ \ \ \ \ \times 
 \left( \begin{array}{cc} \tau_2 & \\ & 1 \end{array} \right) \left( \begin{array}{cc}     \tau_{1}^{-1} +  1  &   \mu_1   \tau_{1}^{-1} + 1 \\  \mu_2   \tau_{1}^{-1} +  1   &    \mu_1 \mu_2   \tau_{1}^{-1} + 1   \end{array} \right)   A'_1 A'_2 A'_3 ... A'_{2n-1} A'_{2n}. $$

Let $k$ be such that $A'_1  A'_2...  A'_{2n} = 1 + E_r k$; more precisely:

\begin{Lemma}  For every $E_r\in \C$ small enough, there exists  a resurgent symbol $k=k(E_r,h)=a_0(E_r)+a_1(E_r)h+a_2(E_r)h^2+$ such that  
$$A'_1 A'_2 ...  A'_{2n} = 1 + E_r k$$ 
and $k$ is holomorphic with respect to $E_r$. 
\end{Lemma}

Remark that the coefficients $a_j(E_r)$ are also holomorphic with respect to $E$, as immediately follows from the iterative procedure of calculating them.

\textsc{Proof of the lemma.} Recall that $A'_j=\frac{\phi_{-}(q_{j+1}-\varepsilon)}{\phi_{-}(q_{j}-\varepsilon)}$, where $\phi_{-}(q,h)$ is the formal solution corresponding to the lower sheet of $p(q)$. This means that the product $A'_1  A'_2...  A'_{2n}$ is the formal monodromy around the loop from $q_0$ to $q_0+1$ of the formal solution corresponding to the second sheet. But for $E_r=0$ this solution can be taken as $\exp[-\frac{f(q)-f(0)}{h}]$ with the trivial monodromy, hence the lemma. 
$\Box$

Introduce a new matrix $G_0$ by $G_0=(1+E_r k)^{-1}G$.

\begin{Lemma} The matrix $G_0$ can be written in the form $$ G_0 \ = \ \left( \begin{array}{cc} 1+E g_{11} & E g_{12} \\ E g_{21} & 1 + E g_{22} \end{array}  \right),$$  
where $g_{ij}$ are resurgent symbols holomorphically dependent on $E_r$. \end{Lemma}

\textsc{Proof.} Easily shown by induction. It is true for $n=1$ and a product of two such matrices is again of this form. $\Box$.

The quantization condition can now be rewritten as $\det (G -   I) = 0$, i.e.
\begin{equation} \left(\frac{1}{1+E_r k}\right)^2 - {\rm Tr}G_0  \frac{1}{1+E_r k} + \det G_0 = 0. \label{QCond} \end{equation}

\subsection{Ingredients of the quantization condition}

In order to solve the quantization condition \eqref{QCond} 
for $E_r$, it is important to understand the determinant and the trace of the matrix $G_0$.

\begin{Lemma} We have $$\det G_0 \ = \ \tau_1^{-1}\tau_2 ... \tau_{2n-1}^{-1}\tau_{2n} (1-\mu_1)(1-\mu_2)...(1-\mu_{2n}) \ = \ 1+E_r d,$$
where $d$ depends holomorphically on $E_r$.
\end{Lemma}

\textsc{Proof.} For the first equality, use multiplicativity of $\det$ and the case $n=1$. The second equality follows from the expressions for  $\tau_j$ and $\mu_j$ established below. $\Box$

To simplify notation, we will calculate $\tau_j$ for $j=1,2$; the similar formulae will hold for other $j$'s.

{\bf Calculation of $\tau_1$.} We know from \ref{ExactMatchingq1} that 
$$c'_1= \frac{2i\sqrt{\pi}f'(q_1-\varepsilon)}{\sqrt{hf''(q_1)}}  e^{\frac{2[f(q_1)-f(q_1-\varepsilon)]}{h}}(1+O(h)+O(E_r)), $$
$$c_2= -E_r \frac{ i\sqrt{\pi h} }{2\sqrt{|f''(q_2)|}f'(q_2-\varepsilon)}  e^{-\frac{2[f(q_2)-f(q_2-\varepsilon)]}{h}}(1+O(h)+O(E_r)), $$
and we are going to calculate that
$$ A_1^{-1} A'_1 \ = \ e^{\frac{-2[f(q_2-\varepsilon)-f(q_1-\varepsilon)]}{h}}\frac{f'(q_2-\varepsilon)}{f'(q_1-\varepsilon)} (1+O(h)+O(E_r)).$$
Together, this will yield
$$\tau_1 = c'_1 c_2 A_1^{-1} A'_1 \ = \ $$ 
$$ \ = \ E_r \frac{\pi}{\sqrt{f''(q_1)|f''(q_2)|}}  e^{-\frac{2[f(q_2)-f(q_1)]}{h}} (1+O(h)+O(E_r)). $$

\begin{figure}[h] \includegraphics{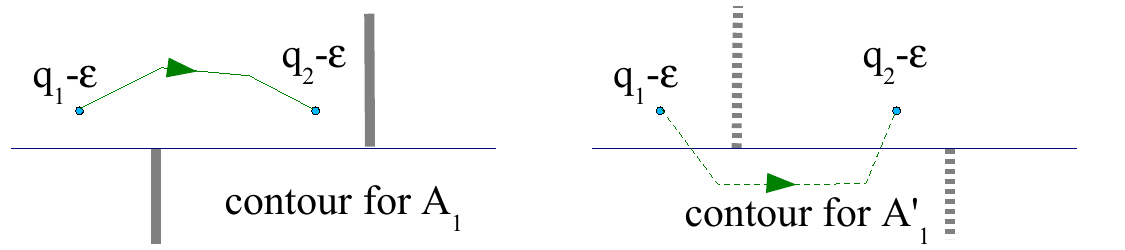} \caption{Integration contours defining $A_1$ and $A'_1$.} \label{nuthesisp20} \end{figure}

This is how $A_1^{-1}A'_1$ is calculated: $A_1$ and $A'_1$ are formal monodromies along the contours shown on figure \ref{nuthesisp20}.

The exponent is obvious. To find the $O(1)$ term, calculate 
$$ \int_{q_1-\varepsilon}^{q_2-\varepsilon} \left[ \frac{f'f''}{2(E-(f')^2)} - \frac{f''}{2i\sqrt{E-(f')^2}}\right] dq, $$ 
where the path of integration is chosen on the first sheet.

\textsc{First summand:} When we calculate the difference of $\int_{q_1-\varepsilon}^{q_2-\varepsilon} \frac{f'f''}{2(E-(f')^2)}dq$ along the paths on the first and on the second sheets, since the function does not change on the first and on the second sheet, we can reduce the question to calculating $\int_{\gamma_1} \frac{f'f''}{2(E-(f')^2)}dq$ and this integral is $-\pi i$. 

For the \textsc{second summand:} Notice that the integral on the first sheet from $q_1-\varepsilon$ to $q_2-\varepsilon$ is 
$$ \int_{q_1-\varepsilon}^{q_2-\varepsilon} \frac{f''}{2i\sqrt{E-(f')^2}} dq \ = \  \int_{q_1-\varepsilon}^{q_2-\varepsilon} \frac{f''}{2f'} dq \ + \ O(E)  \ = \ 
$$
$$ \ = \ \frac{1}{2}\log\left( \frac{f'(q_2-\varepsilon)}{|f'(q_1-\varepsilon)|} \right) + \frac{\pi i}{2} +O(E).$$

A similar calculation works for the second sheet and yields
$$ \int_{q_1-\varepsilon}^{q_2-\varepsilon} \frac{f''}{2i\sqrt{E-(f')^2}} dq 
 \ = \ -\frac{1}{2}\log\left( \frac{f'(q_2-\varepsilon)}{|f'(q_1-\varepsilon)|} \right) + \frac{\pi i}{2} +O(E) ,$$
so we get
$$ A_1^{-1} A'_1 \ = \ e^{-\frac{2[f(q_2-\varepsilon)-f(q_1-\varepsilon)]}{h}}\frac{f'(q_2-\varepsilon)}{f'(q_1-\varepsilon)} (1+O(E)).$$

{\bf Calculation of $\tau_2$} yields analogously 
$$ \tau_2 = c_2 c'_3 A_2 (A'_2)^{-1} =  E_r \frac{ \pi  }{\sqrt{|f''(q_2)|f''(q_3)}}  e^{-\frac{2[f(q_2)-f(q_3)]}{h}}(1+O(h)+O(E_r)).$$

{\bf Remark.} Further terms in the asymptotic expansion of $\tau_j$  can perhaps be calculated similarly to ~\cite[p.82-83]{DP99}.

{\bf Calculation of $\mu_j$.} We defined $\mu_j$ as the products of off-diagonal elements in the connection matrices $C^{(q_j)}$ obtained in subsection \ref{ConnPathNonsingE}. We have for odd $j$: \small
$$  \mu_j \ = \ 1+e^{2\pi i s_{\delta'_j}} \ = \  2\pi i (s_{\delta_j} - \frac{1}{2}) (1+O(E_r)+O(h)) \ = \ E_r \frac{\pi i}{f''(q_j)} (1+O(h)+O(E_r)), $$ \normalsize
and similarly for even $j$:
$$ \mu_j \ = \ 1+e^{2\pi i s_{\delta_j}}  \ = \ E_r \frac{\pi i}{|f''(q_j)|} (1+O(h)+O(E_r)). $$


\section{Resurgent Solutions of a Resurgent Transcendental Equation} \label{ResurgTranscEqu}

The quantization condition will be an equation on $E_r$ whose left hand side can be written as a polynomial in $E_r$, plus a correction that is exponentially small for $h\to 0+$. More precisely, the quantization condition will satisfy the assumptions of the following: 

\begin{Lemma} If we have an ``approximately" polynomial equation on $E_r$ 
\begin{equation} f_n(E_r,h)E_r^n + a_{n-1}f_{n-1}(E_r,h)E_r^{n-1} + ... + a_1 f_1(E_r,h)E_r + a_0f_0(E_r,h) = 0 \label{eq6} \end{equation}
with $|a_{k}|\le \left[C e^{-c/h}\right]^{n-k}$ and $f_k(E_r,h)=1+O(E_r)+O(h)$, then its solutions are of exponential type $\le Ce^{-c/h}$. \end{Lemma}
 
\textsc{Proof} is obvious. $\Box$

\subsection{Newton polygon} \label{NewtonPolygon}

Given an equation of the form \eqref{eq6}, for every term $E_r^j (1+O(E_r)) e^{k/h}$ on its left hand side plot a point with the coordinates $(j,k)$ on the plane and a quadrant with its vertex there and opening in the direction  down and to the right.  The convex hull of the union of these quadrants will be called the {\bf Newton polygon} of our equation. 

\begin{figure}[h]\includegraphics{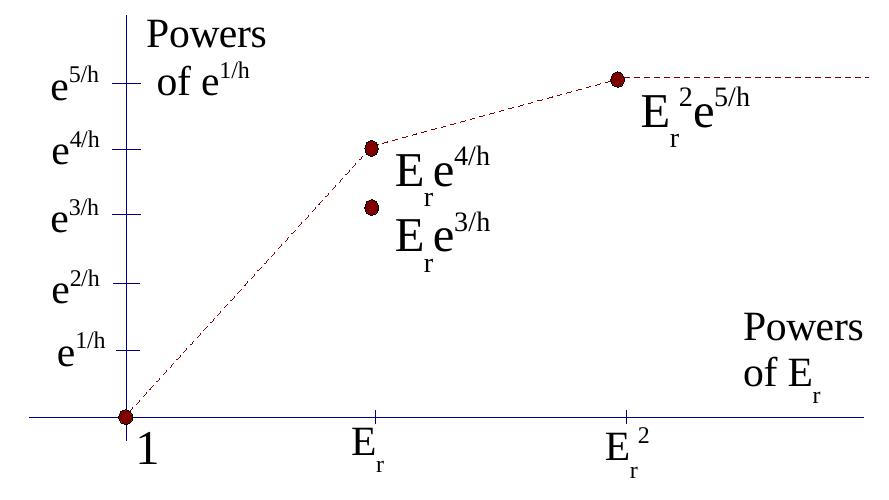} \caption{An example of a Newton polygon.} \label{ResDRWp30}
\end{figure}

E.g., the equation
$$ 3+ (2+h)E_r e^{3/h} + h^2E_re^{4/h} + E_r^2e^{5/h}$$
 produces a Newton polygon shown on figure \ref{ResDRWp30}.
 
If the equation \eqref{eq6} represents a quantization condition for the Witten Laplacian, it will satisfy the following additional properties:

{\it Property 1.} All terms corresponding to the vertices on the boundary of the Newton polygon will be of the form $e^{-c/h}$ times a resurgent (micro)function representable as 
$g(E_r) + (small)$ and that microfuntion has representatives holomorphic with respect to $E_r$.  This follows essentially from the connection formulae and from the properties of a major of a resurgent solution of a differential equation constructed in ~\cite{ShSt}. 

{\it Property 2.} Singular exponents, i.e. the numbers $c$ in the $e^{-c/h}$-factors of the terms of \eqref{eq6},  do not depend on $E_r$, since the classical action $S(q)$ of the Witten Laplacian does not depend on $E_r$.  Therefore we can decompose the LHS of \eqref{eq6} into a sum of resurgent microfunctions corresponding to the hyperasymptotic representation $\sum_i e^{c_i/h} (...)$ with $c_i$ independent of $E_r$.

It will also be important to keep in mind that the construction of a major for substitution of a small resurgent function $r(h)$ for a holomorphic parameter $E$ of $g(E,h)$ implies that the Newton polygon of a composite function can be obtained from the Newton polygons of $g(E,h)$ and of $r(h)$ in exactly the same way as expected from formal manipulations with the formulas.

\subsection{Algebraic equation corresponding to an edge of the Newton polygon}

Consider an equation ${\cal F}(E_r,h)=0$ of type (\ref{eq6}) and its corresponding Newton polygon.  It is clear that if the exponential type of a resurgent symbol $\phi(h)$ is not equal to the slope of an edge of the Newton polygon, then such resurgent symbol cannot be a solution of the equation. 
 
 
Suppose there is an edge of the Newton polygon on the line $y=kx+b$, $k>0$, and the two extreme vertices on this edge are $E^\ell_r e^{\frac{k\ell+b}{h}}$ and $E^{\ell+n} e^{\frac{k[\ell+n]+b}{h}}$. Let us find all resurgent solutions of the equation ${\cal F}(E_r,h)=0$ of exponential type $k$.

\begin{figure}[h]\includegraphics{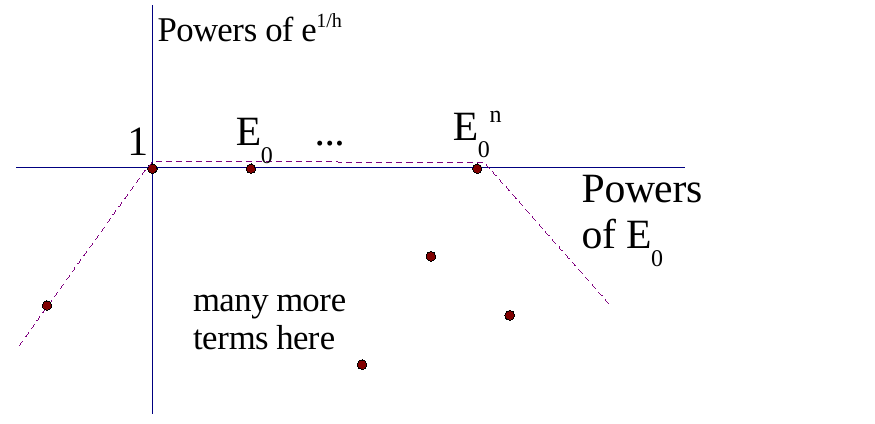} \caption{A Newton polygon after a shearing transformation $E_r=e^{-k/h}E_0$.} \label{RDRWp14}
\end{figure}

A substitution $E_r=e^{-k/h}E_0$ performs a shearing transformation on the Newton polygon. If we plot all the terms of the equation $f_0(E_0,h)=0$, where $f_0(E_0, h)=e^{-b/h} E_0^{-\ell} {\cal F}(e^{-k/h}E_0,h)$,  in the axes corresponding to powers of $E_0$ and of $e^{1/h}$ (figure \ref{RDRWp14}), we will see terms on the horizontal axis between $E_0^0$ and $E_0^{n}$, and all other terms will be below the horizontal axis. We are interested now in finding all resurgent solutions of $f_0(E_0,h)=0$ of zero exponential type; let us show now that in our situation the number of these solutions equals the length $n$ of the edge. 

 Suppose the algebraic equation corresponding to the upper horizontal edge of the polygon on figure \ref{RDRWp14} is of the form 
\begin{equation} A_0+A_1E_0+...+A_nE_0^n \ = \ 0, \label{TopEdge} \end{equation}
where $A_k =a_k + b_k$, where $a_k$ are complex numbers such that the equation $a_0+a_1 x+... + a_n x^n$ has only simple roots and $b_k$ are small resurgent functions. This will be our {\bf nondegeneracy condition} for the superpotential in the Witten Laplacian. 

 Since $A_n$ is invertible, we can assume  $A_n=1$. Let $r$ be a simple root of this equation modulo $O(h)$, then it is resurgent. Indeed, roots of a polynomial equation analytically depend on the coefficients outside of the discriminant locus. Then, if the coefficients are of the form $const(\ne 0)+O(h)$, we can use the theorem on resurgence of composition of a holomorphic function and a small resurgent function (~\cite{G1}) to conclude that roots will be resurgent.
 
The question of whether more degenerate polynomial equations with resurgent coefficients have resurgent roots will be studied elsewhere. Note that the appropriate generality should include resurgent coefficitents given by resurgent power series expansions not simply in $h$, but also in $\ln h$, since this is the form of the tunnel cycle  monodromies $\tau_j$.

\subsubsection{Exponentially small corrections -- the second Newton polygon}

Vi\`ete's formula shows that every one of resurgent roots $r_1,...,r_n$ of (\ref{TopEdge}) has a resurgent inverse; indeed, $r_1...r_n=(-1)^n A_0/A_n$, so $r_j=(-1)^n A_n \times \frac{1}{A_0} \times \prod_{j'\ne j}r_{j'}$ and every factor of this product is resurgent. Now we replace every $E_0^{k}e^{-d/h}$, $k\in \Z$ by $(r+E_1)^k e^{-d/h}$ and get a new Newton polygon with respect to $E_1$ and we will look for solutions of the new equation with respect to $E_1$ which we will require to be exponentially small. That Newton polygon will have $E_1^1$ as its leading term becuase it follows from our assumptions that $r$ is multiplicity one. 

Write our equation in the form
$$ A_0+A_1E_0+...+A_{n-1} E_0^{n-1}+E_0^n + \sum a_{k\ell}E_0^k e^{-c_\ell/h} \ = \ 0 , $$
where $k$ is also allowed to be negative. If $r$ is the root of the polynomial part $A_0+A_1E_0+...+A_nE_0^n=0$, use an ansatz $E_0=r+E_1$. Expanding $(r+E_1)^k$ in powers of $E_1$, obtain
$$ B_1E_1 + ... + B_{n-1}E_1^{n-1}+E_1^n + \sum_{k\ge 0,\ell} e^{-c_\ell/h} b_{kl} E_1^k,$$
where $b_{kl}$ are elementary simple resurgent symbols and 
with $B_1=(const\ne 0)+\phi(h)$ for a small resurgent microfunction $\phi$. 

\begin{figure}[h]\includegraphics{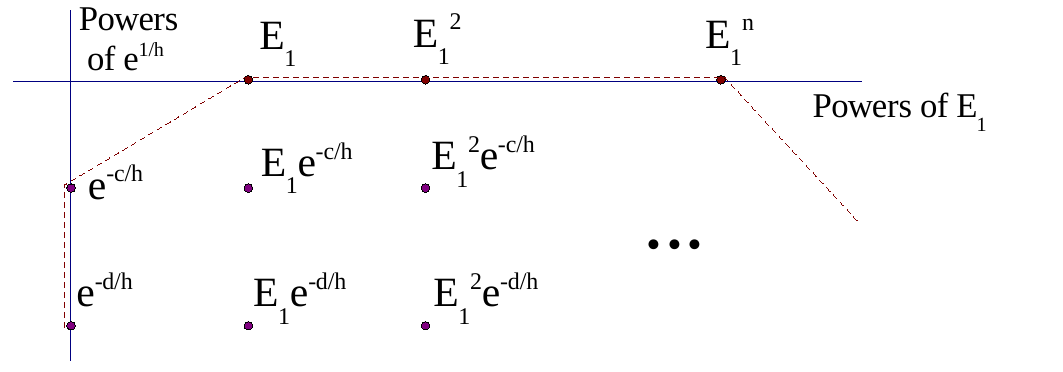} \caption{A Newton polygon in the $E_1$ variable} \label{RDRWp15}
\end{figure}

The Newton polygon for this new equation now looks like the one on figure \ref{RDRWp15}, and it is clear that we get only one exponentially small (or zero) solution.

\begin{figure}[h]\includegraphics{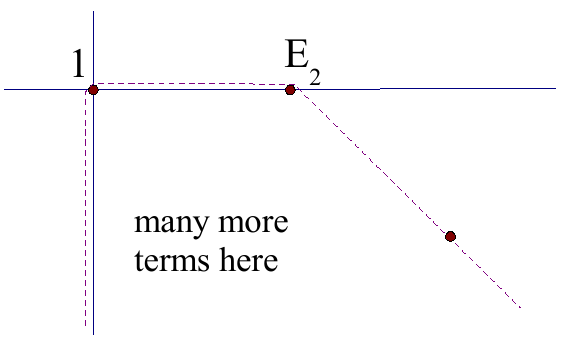} \caption{A Newton polygon in the $E_2$ variable} \label{RDRWp14b}
\end{figure}

In case there are term of degree zero in $E$, there is an upper left edge of the Newton polygon; using its slope, find an ansatz $E_1=e^{-k/h}E_2$, obtain a Newton polygon for $E_2$ with the horizontal edge of length one, see figure \ref{RDRWp14b}.

Let $r_1$ be the solution of the corresponding linear equation on $E_2$, and use the ansatz $E_2=r_1+E_3$ where $E_3$ will be required to stay exponentially small.


\begin{figure}[h]\includegraphics{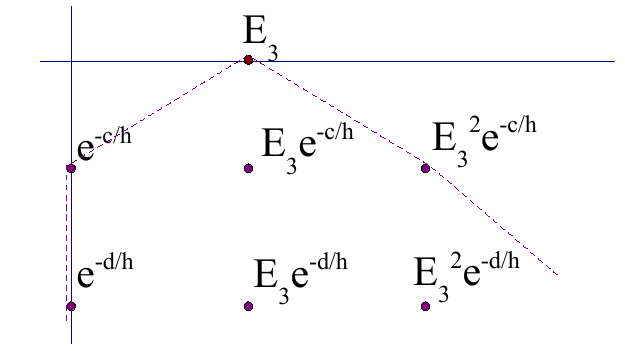} \caption{A Newton polygon in the $E_3$ variable} \label{RDRWp5}
\end{figure}

Again we get a Newton polygon of the same kind. Etc., we keep obtaining exponentially small corrections of smaller and smaller exponential type. 
 On the $j$-th step we are getting solutions modulo ${\cal E}^{-N_j}$, where $N_j$ is a sum of positive numbers $2|f(q_{\nu})-f(q_{\nu'})|$, where $f$ is the function appearing in \eqref{ReducedSchroe} and $q_\nu$ are its real zeros .  Therefore the exponents $N_j$  form a discrete subset of $\R$ and therefore the Mittag-Leffler sum of corrections that we obtain on successive step of the procedure described in this section is a resurgent function.

\subsection{Remarks on justification of the above procedure. }

The terms of the quantization conditions, or of the equation that we have been solving, will be shown in the next section to be polynomials in resurgent symbols $\tau_j$ and $\mu_j$, as will be derived and explained in the next section. There arises the following

{\it Terminology issue:} $\mu(E_r,h)$ is not a function of $h$, it is a resurgent symbol. It would be more appropriate to denote it ${\pmb \mu}(E_r,\zeta)$ and reserve $\mu(E_r,h)$ for ${\cal L}{\bf s}_+ {\pmb \mu}(E_r,\zeta)$. It is shown in ~\cite{G1} that for $E_r$ a number ${\pmb \mu}(E_r,\zeta)$ is holomorphic for $(E_r,\zeta) \in D_\rho \times {\cal S}_\mu$ for some disc $D_\rho \subset \C$ and an endlessly continuable Riemann surface ${\cal S}_\mu$.
The same is true for microfunctions ${\pmb \tau}$ defined on Riemann surfaces ${\cal S}_\tau$.

Further, we know that $\mu$ and $\tau$ are essentially made of (convolution) quotients of microfunctions corresponding to singularities on the Riemann surface of the major of a resurgent solution. As that major can be chosen to analytically depend on $E_r$, so can representatives for ${\pmb \mu}_j$ and ${\pmb \tau}_j$ and also the majors for ${\bf s}_\alpha {\pmb \mu}_j$ and ${\bf s}_\alpha {\pmb \mu}_j$ (cf. ~\cite{G1}). Conclude that the left hand side of our quantization condition $f(E_r,h)=0$, which is a polynomial in ${\pmb \mu}_j$ and ${\pmb \tau}_j$, has a major that holomorphically depends on $E_r$ and defined on the Riemann surface ${}^{\infty}{\cal S}_{\mu_1} * ... *{}^{\infty}{\cal S}_{\tau_{2n}}$.

By construction of a major for $f(E_r(h),h)$ for $E_r(h)$ a small resurgent function carried out in ~\cite{G1} and valid just the same for $E_r(E_1,h)$ for $E_r(E_1,h)$ representable by a major holomorphically dependent on $E_1$, 
one can see that the  Newton polygon for $f(E_r(E_1,h),h)$ (with respect to the powers of $E_1$ now) will be what is expected from the formal manipulation with symbols.


\section{Solving the quantization condition.} \label{Solving}

We think it is helpful at this point to consider \\
{\bf Special case $n=1$,} in which 
$$ G_0 \ = \    
  \left( \begin{array}{cc}     \tau_2 \tau_{1}^{-1} +  \tau_2  &   \tau_2 \mu_1   \tau_{1}^{-1} + \tau_2 \\  \mu_2   \tau_{1}^{-1} +  1   &    \mu_1 \mu_2   \tau_{1}^{-1} + 1   \end{array} \right) ,$$
$${\rm Tr}G_0 = \tau_2 \tau_{1}^{-1} +  \tau_2 +  \mu_1 \mu_2   \tau_{1}^{-1} + 1, \ \ \ \ \det G_0 \ = \ \tau_1^{-1}\tau_2  (1-\mu_1)(1-\mu_2). $$

\begin{figure}[h]\includegraphics{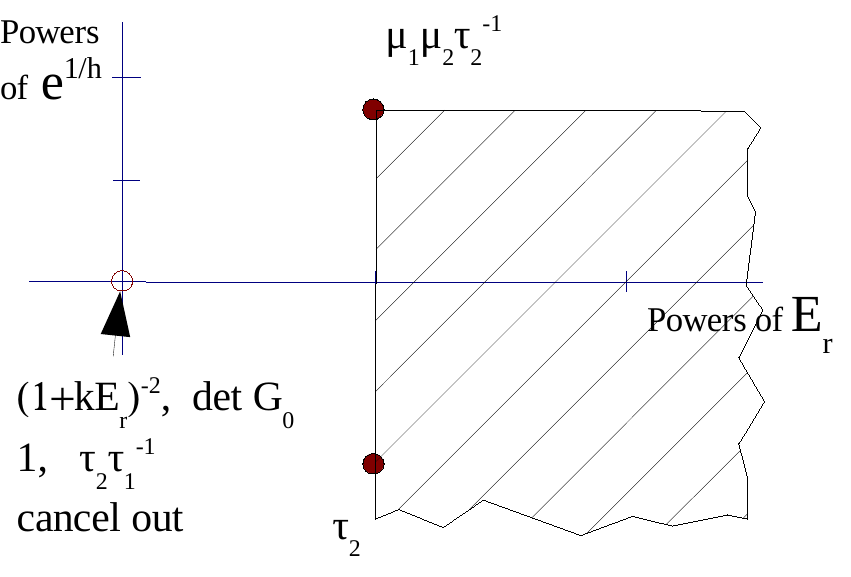} \caption{The Newton polygon in the special case $n=1$.} \label{RDRWp7}
\end{figure}

The Witten Laplacian in this case has an eigenvalue $E_r=0$ corresponding to the eigenfunction $e^{-f(q)/h}$; let us see what it means for the Newton polygon (figure \ref{RDRWp7}) of the quantization condition (\ref{QCond}).


For $E_r=0$ the condition (\ref{QCond}) reduces to 
$$ 1\ - \ \Tr G_0 \ + \ \det G_0 \ = \ 0, $$
or
$$  - \tau_2 - \mu_1\mu_2\tau_1^{-1}  + \tau_1^{-1}\tau_2(-\mu_1 -\mu_2 + \mu_1\mu_2) \ = \ 0, $$
where every term on the left hand side vanishes for $E_r=0$.

We see therefore that all the terms corresponding to the degree (0,0) vertex on figure \ref{RDRWp7} cancel each other, so the Newton polygon will be the hashed subset without edges of positive slope, and hence the quantizaton condition (\ref{QCond}) has no nonzero exponentially small solutions.

\begin{figure} $$ G_0 \ = \ \left( \parbox{6.68cm}{ \includegraphics{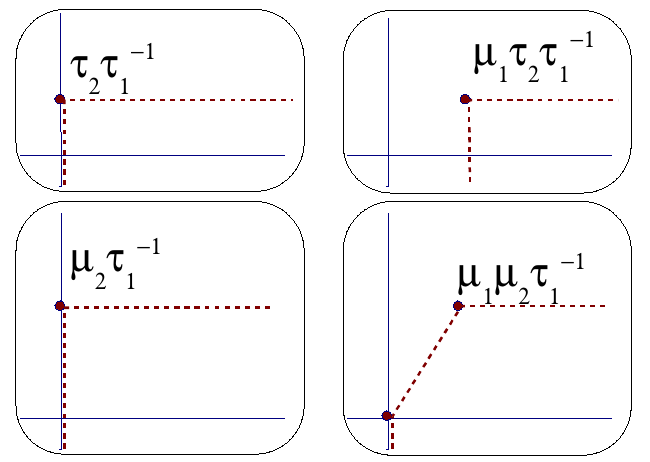}}\right) $$ \caption{Newton polygons for entries of $G_0$ in case $n=1$} \label{NPn1} \end{figure}

\begin{figure} $$  \left( \parbox{10.7cm}{\includegraphics{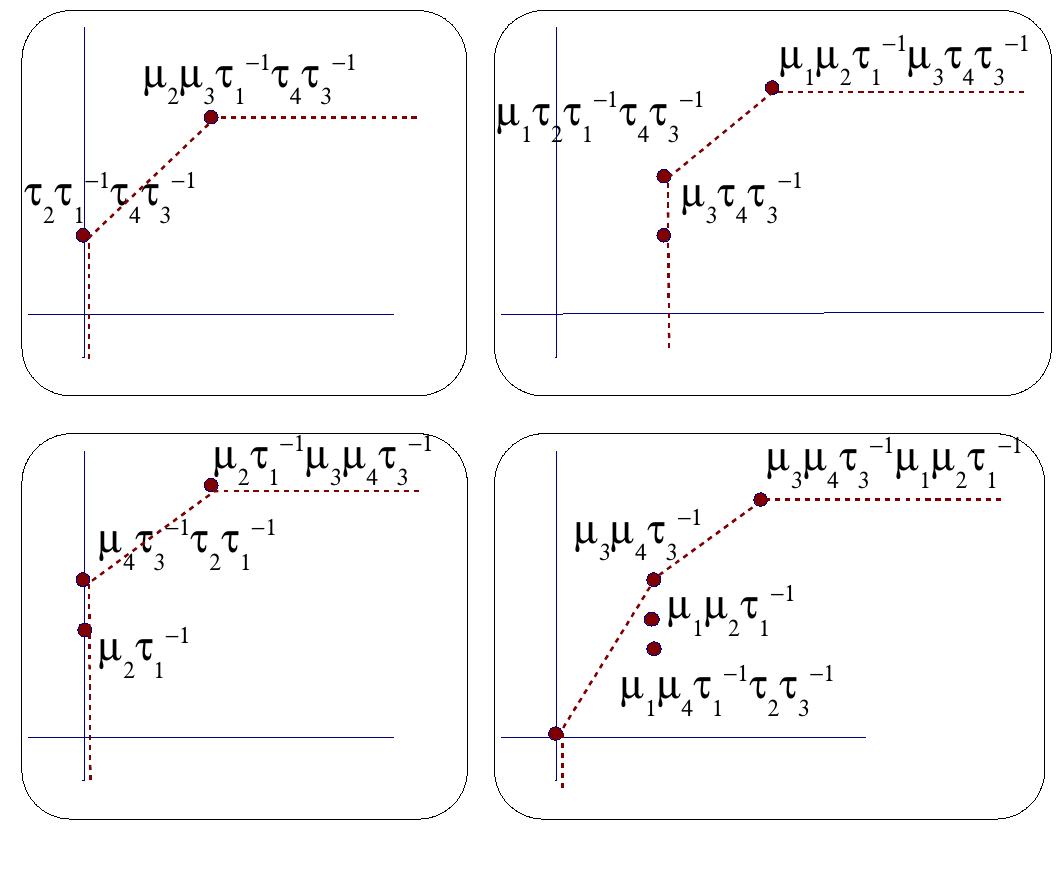}} \right)$$ \caption{Newton polygons for entries of $G_0$ when $n=2$. In case of $(G_0)_{12}$, resp., $(G_0)_{21}$, resp., $(G_0)_{22}$. only terms $\mu_1 \tau_2 \tau_1^{-1}\tau_4\tau_3^{-1}$ and $\mu_3\tau_4\tau_3^{-1}$, resp., $\mu_4\tau_3^{-1}\tau_2\tau_1^{-1}$ and $\mu_2\tau_1^{-1}$, resp., $\mu_3\mu_4\tau_3^{-1}$, $\mu_1\mu_2\tau_1^{-1}$, and $\mu_1\mu_4\tau_1^{-1}\tau_2\tau_3^{-1}$ contrubute to the vertex of the Newton polygon of degree $1$, resp., $0$, resp. $1$, with respect to $E_r$, unless their contributions cancel in which case the corresponding vertex may be absent or lie lower. } \label{NPn2} \end{figure}

{\bf Newton polygon of the quantization condition  in the general case.} For each entry of the matrix $G_0$ one can draw a Newton polygon with respect to powers of $E_r$ and $e^{1/h}$, as prescribed in section \ref{NewtonPolygon}. In case $n=1$ and $n=2$ the Newton polygons are shown on figure \ref{NPn1} and \ref{NPn2}. For drawing these figures as well for the rest of the section we are using that $\tau_j$s are exponentially small for all $j$.

Given a resurgent symbol $g$ dependent on $E_r$ and subject to appropriate conditions, its Newton polygon will have several edges of positive slope; take the right end of the rightmost edge of positive slope and call it the {\bf leading term} of the symbol $g$ and denote it by ${\rm L.T.}g$. Clearly,  ${\rm L.T.} (g_1 g_2) \ = \ {\rm L.T.}g_1  {\rm L.T.}g_2$.

\begin{Lemma} The leading term of the Newton polygon of (the left hand side of) the equation \eqref{QCond} 
is  $\mu_1 \mu_2... \mu_{2n}\tau_1^{-1}\tau_3^{-1}...\tau_{2n-1}^{-1}$ and has degree $n$ in $E_r$.
\end{Lemma}

\textsc{Proof.} We need to look at the terms of $\Tr G_0$ only, because all other terms give contributions of the form $c E^k$ with $c$ of exponential type zero. 

By the $L.T.$ of a matrix we will denote the matrix of leading terms of each of its entries.

One shows by induction on $n$ that 
$$ {\rm L.T.} G_0 \ = \ 
 \mu_1\mu_2...\mu_{2n} \tau_1^{-1} \tau_3^{-1} ... \tau_{2n-1}^{-1} \left( \begin{array}{cc}     \mu_1^{-1}  \mu_{2n}^{-1}  \tau_{2n}  & \mu_{2n}^{-1}\tau_{2n} \\  
\mu_1^{-1}  & 1  \end{array} \right),  $$
from which the statement of the lemma follows. 
$\Box$

This finishes the proof of the main result of this paper, theorem \ref{MainResult}.


\section{Example.} \label{example}

Take as the superpotential 
$$ f = \frac{1}{2\pi}\left[ \sin 2\pi (q + \frac{1}{8}) + \cos 4\pi (q + \frac{1}{8}) \right], $$
$$f' = \cos 2\pi (q + \frac{1}{8}) - 2 \sin 4\pi (q + \frac{1}{8})  .$$ 



The critical points of $f$ are all real in this case:
$$ \begin{array}{cccccc} q_1 = \frac{1}{8} &&& f(q_1) = 0  && f''(q_1)=6\pi,\\  
q_2  = \frac{3}{8} - \frac{1}{2\pi}\arcsin \frac{1}{4} &&& f(q_2)  = \frac{9}{16\pi} && f''(q_2)=-7.5\pi, \\
q_3 = \frac{5}{8} &&& f(q_3) = -\frac{1}{\pi} && f''(q_3)=10\pi, \\ 
q_4 = \frac{7}{8} + \frac{1}{2\pi}\arcsin \frac{1}{4} &&& f(q_4) =  \frac{9}{16\pi} && f''(q_4)=-7.5\pi.
 \end{array}$$


The study of the equation (\ref{QCond}) tells us that the Witten Laplacian corresponding to the function $f$ will have two low-lying eigenvalues: the zero and one nonzero exponentially small eigenvalue that will be expressible in terms of $\mu$s and $\tau$s. The Newton polygon corresponding to the equation (\ref{QCond}) will have a vertex corresponding to the leading term of degree $2$ in $E_r$ that can be obtained by looking at the summand $\mu_1\mu_2\mu_3\mu_4 \tau_1^{-1}\tau_3^{-1}$. Our present task is to find the vertex of the Newton polygon corresponding to term of degree $1$ with respect to $E_r$.

We get (using the formulas for $\tau_k$)
$$\begin{array}{ccc} \tau_1 \approx e^{-\frac{9}{8\pi h}}\frac{E_r}{\sqrt{6\times 7.5}} & & \tau_2 \approx e^{-\frac{25}{8\pi h}}\frac{E_r}{\sqrt{10\times 7.5}} \\ 
\tau_3 \approx e^{-\frac{25}{8\pi h}} \frac{E_r}{\sqrt{10\times 7.5}}& & \tau_4 \approx e^{-\frac{9}{8\pi h}}\frac{E_r}{\sqrt{6\times 7.5}} 
\end{array}$$

$$\begin{array}{ccc} \mu_1 \approx i\frac{E_r}{6} && \mu_2 \approx i\frac{E_r}{7.5} \\
\mu_3 \approx i\frac{E_r}{10} && \mu_4 \approx i\frac{E_r}{7.5} 
\end{array}$$

We have seen in the picture that the there are four terms in the equation (\ref{QCond}) that can produce the vertex of $\deg_{E_r} = 1$, and they are:
$$\mu_2 \mu_3 \tau_1^{-1} \tau_4 \tau_3^{-1} \ \approx \ 
-E_r \frac{1}{\sqrt{10\times 7.5}}e^{\frac{25}{8\pi h}}, $$
$$\mu_3 \mu_4 \tau_3^{-1}  \ \approx  \ 
-E_r \frac{1}{\sqrt{10\times 7.5}}e^{\frac{25}{8\pi h}},$$
$$\mu_1 \mu_2 \tau_1^{-1}  \ \approx \ -E_r \frac{1}{\sqrt{6\times 7.5}} e^{\frac{9}{8\pi h}},$$
$$ \mu_1 \mu_4  \tau_1^{-1}  \tau_2 \tau_3^{-1} \ \approx \ -E_r\frac{1}{\sqrt{6\times 7.5}} e^{\frac{9}{8\pi h}}. $$
Therefore the corresponding vertex comes as a sum of contibutions of the first two summands and is located at $E_r e^{\frac{17}{8\pi h}}$.

The leading term in the Newton polygon is
$$ \mu_1\mu_2\mu_3\mu_4 \tau_1^{-1} \tau_3^{-1} \approx E_r^2 \frac{1}{\sqrt{6\times 7.5\times 10 \times 7.5}}e^{\frac{25+9}{8\pi h}},$$
Hence $$E_r \approx  2\frac{\frac{1}{\sqrt{10\times 7.5}}e^{\frac{25}{8\pi h}}}{\frac{1}{\sqrt{6\times 7.5\times 10 \times 7.5}}e^{\frac{25+9}{8\pi h}}}  \approx 2\sqrt{6\times 7.5}e^{-\frac{9}{8\pi h}}.$$

{\it Remark 1.} In a more general case, since $\mu_j\in i\R_+$ and $\tau_j\in\R_+$  for $E>0$, $h>0$, conclude that the term in the numerator cannot cancel

{\it Remark 2.} The result is clearly what one would expect from ~\cite{HeKlNi}, but our case does not satisfy their nondegeneracy conditions.

{\bf \large Acknowledgements } 

The author would like to thank his advisor Boris Tsygan for a wonderful graduate experience and his dissertation committee members Dmitry Tamarkin and Jared Wunsch for their constant support in his study and research. 

Valuable comments and suggestions were also made by M.Aldi, J.E.An\-der\-sen, K.Burns, K.Costello, E.Delabaere, S. Garoufalidis, E.Getzler, B.Helf\-fer, A.Karabegov, S.Koshkin, Yu.I.Manin, G.Masbaum,  D.Nadler, W.Richter, K.Vilonen, F.Wang, E.Zaslow, M.Zworski, and by anonymous referees.

This work was partially supported by the NSF grant DMS-0306624 and the Northwestern University WCAS Dissertation and Research Fellowship.


\end{document}